%% file: main.tex
\makeatletter
\def\cl@chapter{}
\makeatother

\documentclass[natbib, draft]{svjour3}

\usepackage{mathptmx, newtxtext}
\usepackage[T1]{fontenc}

\usepackage[english]{babel}
\journalname{Empirical Software Engineering}

\usepackage[utf8]{inputenc}
\usepackage{siunitx}

\usepackage[final]{graphicx}
\usepackage{xcolor}

\usepackage{booktabs}
\usepackage{multirow}
\usepackage{nicematrix}

\usepackage{amsmath}
\usepackage{amsfonts}
\usepackage{amssymb}
\usepackage{centernot}
\usepackage{mathtools}

\usepackage[final]{hyperref}
\usepackage[nameinlink]{cleveref}
\crefname{result}{Simulation~Result}{Simulation~Results}
\Crefname{result}{Simulation~Result}{Simulation~Results}

\usepackage{subcaption}
\usepackage{tikz}
\usepackage{ifthen}
\usepackage{pgfplots}
\usetikzlibrary{automata, positioning, arrows, decorations.pathreplacing, calc, chains, shapes.misc, backgrounds, fit, fadings, matrix, trees, intersections, pgfplots.fillbetween, tikzmark, patterns, patterns.meta}

\pgfdeclarelayer{background}
\pgfsetlayers{background, main}

\input{tikz-hypergraph}

\definecolor{blue}{RGB}{0, 114, 178}
\definecolor{cyan}{RGB}{86, 180, 233}
\definecolor{green}{RGB}{0, 158, 115}
\definecolor{yellow}{RGB}{240, 228, 66}
\definecolor{orange}{RGB}{230,159,0}
\definecolor{red}{RGB}{213, 94, 0}
\definecolor{purple}{RGB}{204, 121, 167}

\colorlet{e1_color}{blue}
\colorlet{e2_color}{red}
\colorlet{e3_color}{green}
\colorlet{e4_color}{purple}

\colorlet{trivago_color}{red}
\colorlet{spotify_color}{green}
\colorlet{microsoft_color}{blue}

\tikzset{brace/.style={decorate, decoration={brace}},
	brace mirrored/.style={decorate, decoration={brace,mirror}},
}

\newcounter{brace}
\pgfplotsset{
	compat=newest,
	every axis plot/.append style={very thick},
	ecdf axis/.style={
		width=\textwidth,
		height=0.66\textwidth, 
		enlargelimits=false,
		clip=false,
		grid=both,
		ymax=1.0,
		ymin=0.0, 
		ytick = {0, 0.1, 0.2, 0.3, 0.4, 0.5, 0.6, 0.7, 0.8, 0.9, 1.0},
		legend cell align={left},
		xtick align=outside,
		ytick align=outside,
		every tick/.style={black},
		axis x line*=bottom,
		axis y line*=left,
	}
}

\newcommand{\menu}[2]{\mbox{#1 $\rightarrow$ #2}}

\usepackage[edges]{forest}

\usepackage[framemethod=TikZ]{mdframed}

\newcounter{result}[section]
\renewcommand{\theresult}{\arabic{result}}

\newenvironment{result}[1]{
	\refstepcounter{result}
	\begin{mdframed}[
		frametitle={\colorbox{white}{\space Simulation~Result \theresult\space}},
		innertopmargin=0pt,
		frametitleaboveskip=-\ht\strutbox,
		frametitlealignment=\raggedright,
		nobreak=true,
		skipabove=10pt,
		skipbelow=10pt,
	]%
	\label{#1}}{\end{mdframed}}

\newcommand{\FigureLink}{%
    \tikz[x=1.2ex, y=1.2ex, baseline=-0.05ex]{%
        \begin{scope}[x=1ex, y=1ex]
            \clip (-0.1,-0.1) 
                --++ (-0, 1.2) 
                --++ (0.6, 0) 
                --++ (0, -0.6) 
                --++ (0.6, 0) 
                --++ (0, -1);
            \path[draw, 
                line width = 0.5, 
                rounded corners=0.5] 
                (0,0) rectangle (1,1);
        \end{scope}
        \path[draw, line width = 0.5] (0.5, 0.5) 
            -- (1, 1);
        \path[draw, line width = 0.5] (0.6, 1) 
            -- (1, 1) -- (1, 0.6);
	}
}

\newcommand{\link}[1]{\begin{flushright}\FigureLink~\Cref{#1}\end{flushright}}

\begin{document}

\title{The upper bound of information diffusion in code review}
\author{Michael~Dorner \and Daniel~Mendez \and Krzysztof~Wnuk \and Ehsan~Zabardast \and Jacek~Czerwonka}
\institute{
Michael Dorner \and Krzysztof Wnuk \and Ehsan Zabardast \at Blekinge Institute of Technology, Karlskrona, Sweden 
\and
Daniel Mendez \at Blekinge Institute of Technology, Karlskrona, Sweden
and Fortiss, München, Germany
\and
Jacek Czerwonka \at Microsoft, Seattle, USA}

\maketitle

\begin{abstract} 
\leavevmode\par
\noindent\textbf{Background: } %
Code review, the discussion around a code change among humans, forms a communication network that enables its participants to exchange and spread information. Although reported by qualitative studies, our understanding of the capability of code review as a communication network is still limited. 

\noindent\textbf{Objective: } %
In this article, we report on a first step towards understanding and evaluating the capability of code review as a communication network by quantifying how fast and how far information can spread through code review: the upper bound of information diffusion in code review.

\noindent\textbf{Method: } %
In an in-silico experiment, we simulate an artificial information diffusion within large (Microsoft), mid-sized (Spotify), and small code review systems (Trivago) modelled as communication networks. We then measure the minimal topological and temporal distances between the participants to quantify how far and how fast information can spread in code review. 

\noindent\textbf{Results: } %
An average code review participants in the small and mid-sized code review systems can spread information to between \qty{72}{\percent} and \qty{85}{\percent} of all code review participants within four weeks independently of network size and tooling; for the large code review systems, we found an absolute boundary of about \num{11000}  reachable participants. On average (median), information can spread between two participants in code review in less than five hops and less than five days.

\noindent\textbf{Conclusion: } %
We found evidence that the communication network emerging from code review scales well and spreads information fast and broadly, corroborating the findings of prior qualitative work. The study lays the foundation for understanding and improving code review as a communication network. 
\end{abstract}

\section{Introduction}

Modern software systems are often too large, too complex, and evolve too fast for a single developer to oversee all parts of the software and, thus, to understand all implications of a change. Therefore, most software projects rely on code review to foster discussions on changes and their impacts before they are merged into the code bases to assure and maintain the quality of the software system. All available and required information about a change can become evident, transparent, and explicit through those discussions and can be shared among the participants. The discussion participants can leverage this information for their own work and pass it on in the following code reviews; the information diffuses through the communication network that emerges from code review.

Five qualitative studies have so far reported on the transition of code review from a more waterfall-like procedure used for detecting bugs in formal, heavyweight code inspections as done in the 1980s towards a more informal, tool-supported, and lightweight communication network for developers to provide and receive relevant and context-specific information for the code change \citep{Bacchelli2013, Rigby2013, Baum20161, Bosu2017, Sadowski2018}. 

Available qualitative studies strengthen already our confidence in the motivation for and expectation towards modern code review as a communication network. However, there is still little to no research that has quantified and measured the actual capability of code review as a communication network. In this article, we report on our experiment results that complement and corroborate those five available qualitative studies.

The objective of our study is to make a first step towards better understanding and evaluating code review as a communication network by quantifying how far and how fast information can spread among the participants in code review. 

In detail, we set out the following two research questions to answer:

\begin{center}
\begin{tabular}{rl}
\textbf{RQ~1} & How far can information spread through code review?\\
\textbf{RQ~2} & How fast can information spread through code review?
\end{tabular}
\end{center}

We address those two research questions in an \emph{in-silico} experiment that simulates an artificial information diffusion in code review networks at three industry cases of different sizes and different code review tools: Microsoft, Spotify, and Trivago. The simulated information diffusion within the communication networks identifies all minimal time-respecting paths reflecting information diffusing through the communication network under best-case assumptions. The participants along those minimal time-respecting paths describe how far information can spread among code review participants (RQ~1), and the minimal topological and temporal distances between participants describe how fast information spreads (RQ~2). Both measures together allow us to better understand the upper bound of information diffusion in code review. 

The main contribution of our study is an in-silico experiment to simulate information diffusion within three industrial code review systems to provide a quantitative assessment of code review as a communication network under best-case assumptions. Beyond this main contribution, we also synthesize qualitative findings from prior work regarding the expectations and motivations towards code review as motivation for our work and provide an extensive and thoroughly engineered replication package.

For this article, we define code review as the informal and asynchronous discussion around a code change among humans. This means older results from formal code inspections and pair programming as an informal but synchronous discussion around a code change among usually two developers are beyond the scope of our study. 

In our study, we focus on code review in an industrial context. Although code review is nearly omnipresent in open source as well and the results are not necessarily contradicting, we strongly believe that results and findings from open source, such as \cite{Rigby2011, Pascarella2018} and \cite{Rigby2008}, are not directly transferable to industrial settings without further considerations. The mechanics and incentives in open source differ, and so do the organizational structure, liability, and commitment \citep{Barcomb2020}. 

The remainder of this paper is structured as follows: In \Cref{sec:background}, we provide an overview of the state of the art on the expectation towards code review, measuring information exchange in code review, information diffusion, and simulation as empirical research method. \Cref{sec:experimental_design} describes our simulation model (\Cref{sec:simulation_model}) and its empirical parametrization (\Cref{sec:parametrization}) in detail. After we report and discuss the simulation results in \Cref{sec:results,sec:discussion}, we discuss the limitations of our work in \Cref{sec:limitations} and close our article with a conclusion and outlook on future work in \Cref{sec:conclusion}.

\section{Background}\label{sec:background}

\cite{Badampudi2023} identified different research themes on code review in a large systematic mapping study where the authors analyzed 244 primary studies until 2021 (inclusive). They further assessed the practitioners' perceptions on the relevance of those code review research themes through a survey of 25 practitioners. 68\% of the practitioners from the survey mentioned the importance of conducting research on a more differentiated view of improvements through code review going beyond finding defects. Our research aims to fill that gap. 

In the following, we elaborate on background and related work with special attention to synthesizing existing qualitative studies. We explore, in particular, the expectations towards code review in industry before laying the foundation for our simulation study by discussing measurements in information exchange, information diffusion, and, more generally, simulations as an empirical research method.

\subsection{Expectations Towards Code Review in Industry}\label{sec:expectations}

Although \cite{Nazir2020} report on preliminary results of a systematic mapping study on the expected benefits of code review, we could not reconstruct how and why the proposed themes---in particular those for knowledge sharing---map the referenced work. Moreover, the study does not distinguish between the expectations towards code review in an open-source and an industrial setting which, as we argued, are not necessarily comparable due to the differences in incentives, organizational structure, liability, and commitment. 

In this section, we, therefore, concentrate on discussing and synthesizing five qualitative studies which have investigated the motivations and expectations towards code review in an industrial context: \cite{Bacchelli2013, Baum20161, Bosu2017, Sadowski2018, Cunha2021}. 

\Cref{tab_expectations} summarizes the findings among the five prior qualitative work and their definition. 

\begin{table}
\centering
\caption{Expectations towards code review reported in \cite{Bacchelli2013, Baum20161, Bosu2017, Sadowski2018, Cunha2021}.}
\label{tab_expectations}
\begin{tabular}{@{}lp{2.8cm}p{5cm}@{}} 
\toprule
Identifier & {Expectation} & {Definition} \\ \midrule
\menu{\cite{Bacchelli2013}}{1} & Finding defects & without explicit definition, presumably comments or changes on correctness or defects in alignment with the \menu{\cite{Bacchelli2013}}{2} \\
\menu{\cite{Bacchelli2013}}{2} & Code improvements & ``Comments or changes about code in terms of readability, commenting, consistency, dead code removal, etc., [without comments or changes] on correctness or defects'' \\
\menu{\cite{Bacchelli2013}}{3} & Alternative solutions & ``Changes and comments on improving the submitted code by adopting an idea that leads to a better implementation'' \\
\menu{\cite{Bacchelli2013}}{4} & Team awareness and transparency & without explicit definition, improved information flow across team boundaries \\
\menu{\cite{Bacchelli2013}}{5} & Share code ownership & without explicit definition \\
\menu{\cite{Bacchelli2013}}{6} & Knowledge sharing (or learning) & without explicit definition \\
\menu{\cite{Baum20161}}{1} & Finding defects & Improved external code quality \\
\menu{\cite{Baum20161}}{2} & Better code quality & Improved internal code quality \\
\menu{\cite{Baum20161}}{3} & Finding better solutions & Finding new or better solutions \\
\menu{\cite{Baum20161}}{4} & Sense of mutual responsibility & Improved collective code ownership and solidarity \\
\menu{\cite{Baum20161}}{5} & Compliance to QA guidelines & Compliance to standards or regulatory norms \\
\menu{\cite{Baum20161}}{6} & Learning (reviewer) & Learning for the author of the code change \\
\menu{\cite{Baum20161}}{7} & Learning (author) & Learning for the reviewer of the code change \\
\menu{\cite{Bosu2017}}{1} & Maintainability & ``legibility, testability, adherence to style guidelines, adherence to application integrity, and conformance to project requirements'' \\
\menu{\cite{Bosu2017}}{2} & Knowledge sharing & ``Code review facilitates multiple types of knowledge sharing.'', ``Code review interactions help both authors and reviewers learn how to solve problems [...]''\\
\menu{\cite{Bosu2017}}{3} & Functional defects & Eliminating ``logical errors, corner cases, security issues, or general incompatibility problems'' \\
\menu{\cite{Bosu2017}}{4} & Community building & without further definition \\
\menu{\cite{Bosu2017}}{5} & Minor errors, typos & without further definition \\
\menu{\cite{Sadowski2018}}{1} & Accident prevention & Avoiding the introduction of bugs, defects, or other quality-related issues \\
\menu{\cite{Sadowski2018}}{2} & Gatekeeping & ``Establishment or maintenance of boundaries around source code, design choices, or other artifacts'' \\
\menu{\cite{Sadowski2018}}{3} & Maintaining norms & ``Organization preference for a discretionary choice, e.g., formatting or API usage patterns'' \\ 
\menu{\cite{Sadowski2018}}{4} & Education & Learning and teaching from code review \\ 
\menu{\cite{Cunha2021}}{1} & Code-related aspects & without explicit definition \\ 
\menu{\cite{Cunha2021}}{2} & Share knowledge on the team or project \& team & without explicit definition\\ 
\menu{\cite{Cunha2021}}{3} & Sharing knowledge between different seniority levels or roles & without explicit definition \\
\bottomrule
\end{tabular}
\end{table} 

In a mixed-method approach, \cite{Bacchelli2013} explored the expectations, outcomes, and challenges of modern code review at Microsoft. From analyzing code review comments and interviews with developers and managers at Microsoft, this seminal study identifies ten different motivations for and expectations towards code review and concludes that although finding defects is a key motivation for code review, only a small portion of the code review comments were defect-related and ``mainly cover small, low-level issues''. The six motivations listed in \Cref{tab_expectations} are discussed in detail; the other four motivations are not discussed further in the paper. Not all motivations are explicitly defined and, in our opinion, are not necessarily mutually exclusive. In particular, exploring the relationship between knowledge sharing and the other expectations further was not in the scope of the study. The study thus recommends studying the socio-technical effects of and investigating if and how learning increases as a result of code review. 

The study by \cite{Baum20161} reports on ten effects (seven desired and three undesired effects) of code review using grounded theory as part of an interview study with 24 software engineering professionals from 19 companies. The study reports seven findings as desired code review effects. Thereby, the study implicitly confirms the reported motivations from \cite{Bacchelli2013} although the authors do not discuss the relation to existing evidence explicitly. A frequency or weighting of the findings is not reported and we may, thus, assume an arbitrary ordering. Through the explicit separation between learning for the author and learning for the reviewer, the study also finds a mutual knowledge transfer, a mutual information exchange, which also supports the notion of bidirectional knowledge transfer as stated in \cite{Bacchelli2013}. 

In two surveys among developers from both an industrial and an open-source context, \cite{Bosu2017} contrasted industrial code review at Microsoft with code review at different open-source projects. The primary motivations reported are maintainability, knowledge-sharing, functional defects, community building, minor errors, and others. They found a significant difference in the primary purposes of code review (``RQ~1: Why are code review important'') between those two contexts: Open-source developers focus more on knowledge-sharing while developers in open source reported maintainability as a primary expectation towards code review. Eliminating functional defects was only the third most important reason for code reviews in both surveys, which further corroborates the findings by \cite{Bacchelli2013}. However, both studies, \cite{Bacchelli2013} and \cite{Bosu2017}, surveyed the same company: Microsoft. This limits more generalizable conclusions from the authors' findings on our side. Although we also rely on the code review system from Microsoft, we added two further industrial code review systems, Trivago and Spotify. This allows us to broaden our perspective on code review in industry. 

In the context of a mixed-methods study, \cite{Sadowski2018} conducted an interview study to investigate the motivations for code review at Google. In more detail, the authors conducted interviews with 12 employees working for Google from one month to ten years (with a median of five years). Four key themes emerged: education (learning and teaching from code review), \emph{maintaining norms} (organization preference for a discretionary choice, e.g., formatting or API usage patterns), \emph{gatekeeping} (establishment or maintenance of boundaries around source code, design choices, or other artifacts\footnote{Upon our request during this study, the authors clarified that gatekeeping refers to requiring a code review from a project owner in order to check in code within a project from someone outside or requiring someone with certification in a particular language to review some code in that language.}), and \emph{accident prevention} (reducing introduction of bugs, defects, or other quality-related issues). The authors explain that these expectations can map over those found previously at Microsoft in \cite{Bacchelli2013} and \cite{Bosu2017}. Unfortunately, the exact mapping is not presented in the article. The authors emphasize that the main focus at Google, as explained by their participants, is on education as well as code readability and understandability. Why, as stated in the manuscript, this focus contradicts the finding by \cite{Rigby2013} that code review has changed from a defect-finding activity to a group problem-solving activity is not discussed further. 

\cite{Cunha2021} report a qualitative survey with 106 practitioners regarding their experiences with modern code review. The paper presents its findings around three codes from the open coding: ``code-related aspects'', ``share knowledge on the team or project'', and ``share knowledge between different seniority levels and roles''. Although details on the practitioners' affiliations are not reported, the surveyed practitioners are affiliated with companies based in Brazil and (in a ``smaller'' yet unreported proportion) in South Africa, Sweden, Ireland, Spain, and France. This broadens the geographical perspective on the expectations towards code review since \cite{Baum20161} surveyed companies based in Germany, the Czech Republic, and the USA, \cite{Bacchelli2013}, \cite{Bosu2017}, \cite{Sadowski2018} report on companies based in the USA (Microsoft and Google). 

All five studies have in common that the use of the terms knowledge sharing, transfer, spreading, or learning is neither consistent among (and partially even within) those prior works nor thoroughly defined. This is likely rooted in the complex nature of knowledge and the different epistemological stances. Furthermore, it remains unclear to what extent knowledge transfer differs from all other expectations. For example, knowledge must be transferred when an alternative solution is proposed or a defect is made explicit through a code review. 

For the synthesis from prior work on the expectations towards code review, we made the following decisions:
\begin{itemize}
\item {\itshape Information over knowledge}---We consistently use \emph{information} instead of knowledge for the synthesis of the prior work and throughout this study. We, thereby, concur with \cite{Pascarella2018}. Although not equivalent, information encodes knowledge since knowledge is the meaning that may be derived from information through interpretation. This means that we may see information as a superset of knowledge. Hence, not all information is necessarily knowledge, but all knowledge is information. This allows us to subsume different stances, definitions, and notions of knowledge without an epistemological reflection upon the various definitions of knowledge. Furthermore, we can refrain from delineating the notion of knowledge from the notion of truth, the latter being too often an inherent connotation of knowledge. We may well postulate that not everything communicated is true. Opinions, expectations, misunderstandings, or best guesses are also part of any engineering and development process and do not meet knowledge and, consequently, truth by all definitions. 
\item {\itshape Information exchange, sharing, spreading, or transfer}---We consider information sharing, spreading, or transfer as synonyms for communication, which is the exchange of information. We will discuss and derive our definition of communication, the exchange of information, in \Cref{sec:conceptual_model} in detail. 
\item {\itshape Improvements over benefits}---The term \emph{benefit} implies that there is a positive outcome from a particular action, decision, or situation---from code review. Since code review does not happen in the void, we prefer the term \emph{improvement} to emphasize the context of code review.
\end{itemize}

\begin{figure}
\centering
\begin{forest}
forked edges,
for tree={
    calign=last,
},        
where level=0{l sep=0.5cm, s sep=2em}{},
where level=1{minimum width=14em}{},
where level<=4{draw}{},
where level>=1{folder, grow'=0, l sep=1em, s sep=0.5em}{},
findings/.style={
	draw=none,
	minimum height=1em,
},
[, phantom
	[{Information exchange}, minimum height=2em, name=ie
		[, name=ie1, findings]
		[, name=ie2, findings]
		[, name=ie3, findings]
		[, name=ie4, findings]
		[, name=ie5, findings]
		[, name=ie6, findings]
		[, name=ie7, findings]
	]
	[{Improvements}, name=improvements, minimum height=2em, 
		[Code, tier=0, minimum height=1em
			[{Functional}, tier=1, s sep=0.5em
				[, name=functional1, findings]
				[, name=functional2, findings]
				[, name=functional3, findings]
				[, name=functional4, findings]
			]
			[Non-functional, tier=1, s sep=0.5em
				[Higher maintainability, tier=2, s sep=0.5em
					[, name=maintainability1, findings]
					[, name=maintainability2, findings]
					[, name=maintainability3, findings]
					[, name=maintainability4, findings]
				]
				[Alternative solutions, tier=2
					[, name=alternativesolution1, findings]
					[, name=alternativesolution2, findings]
					[, name=alternativesolution3, findings]
				]
				[{Compliance with norms \& regulations}, tier=2, s sep=0.5em
					[, name=compliance1, findings]
					[, name=compliance2, findings]
					[, name=compliance3, findings]
					[, name=compliance4, findings]
				]
			]
		]
		[Collaboration, tier=0, s sep=0.5em, minimum height=1em
			[, name=collaboration1, findings]
			[, name=collaboration2, findings]
			[, name=collaboration3, findings]
			[, name=collaboration4, findings]
		]
	]
]
\node[above=1em] (cause) at (ie.north) {Cause};
\node[above=1em] (effect) at (improvements.north) {Effect};
\draw[-stealth] (cause.east) -- (effect.west);
\node[anchor=west] at (functional1.west) {\menu{\cite{Bacchelli2013}}{1}};
\node[anchor=west] at (functional2.west) {\menu{\cite{Baum20161}}{1}};
\node[anchor=west] at (functional3.west) {\menu{\cite{Sadowski2018}}{1}};
\node[anchor=west] at (functional4.west) {\menu{\cite{Bosu2017}}{3}};
\node[anchor=west] at (alternativesolution1.west) {\menu{\cite{Bacchelli2013}}{3}};
\node[anchor=west] at (alternativesolution2.west) {\menu{\cite{Baum20161}}{3}};
\node[anchor=west] at (alternativesolution3.west) {\menu{\cite{Cunha2021}}{1}};
\node[anchor=west] at (maintainability1.west) {\menu{\cite{Bacchelli2013}}{2}};
\node[anchor=west] at (maintainability2.west) {\menu{\cite{Baum20161}}{2}};
\node[anchor=west] at (maintainability3.west) {\menu{\cite{Cunha2021}}{1}};
\node[anchor=west] at (maintainability4.west) {\menu{\cite{Bosu2017}}{1}};
\node[anchor=west] at (compliance1.west) {\menu{\cite{Baum20161}}{5}};
\node[anchor=west] at (compliance2.west) {\menu{\cite{Sadowski2018}}{2}};
\node[anchor=west] at (compliance3.west) {\menu{\cite{Sadowski2018}}{3}};
\node[anchor=west] at (compliance4.west) {\menu{\cite{Cunha2021}}{1}};
\node[anchor=west] at (collaboration1.west) {\menu{\cite{Bacchelli2013}}{4}};
\node[anchor=west] at (collaboration2.west) {\menu{\cite{Bacchelli2013}}{5}};
\node[anchor=west] at (collaboration3.west) {\menu{\cite{Baum20161}}{4}};
\node[anchor=west] at (collaboration4.west) {\menu{\cite{Bosu2017}}{4}};
\node[anchor=west] at (ie1.west) {\menu{\cite{Bacchelli2013}}{6}};
\node[anchor=west] at (ie2.west) {\menu{\cite{Baum20161}}{6}};
\node[anchor=west] at (ie3.west) {\menu{\cite{Baum20161}}{7}};
\node[anchor=west] at (ie4.west) {\menu{\cite{Sadowski2018}}{4}};
\node[anchor=west] at (ie5.west) {\menu{\cite{Cunha2021}}{2}};
\node[anchor=west] at (ie6.west) {\menu{\cite{Cunha2021}}{3}};
\node[anchor=west] at (ie7.west) {\menu{\cite{Bosu2017}}{2}};
\begin{scope}[on background layer]
\node[draw, dashed, rounded corners, fit=(cause)(ie)(ie7), inner sep=1em, label={below:Focus of our study}] {};
\end{scope}
\end{forest}

\caption{Synthesis of expectations towards code review reported by \cite{Bacchelli2013, Baum20161, Bosu2017, Sadowski2018, Cunha2021}: We consider information exchange the cause for the expected improvements from code review. In this study, we aim to provide a quantitative counterpart on information exchange as qualitatively-reported expectation towards code review.
}
\label{fig:synthesis}
\end{figure}
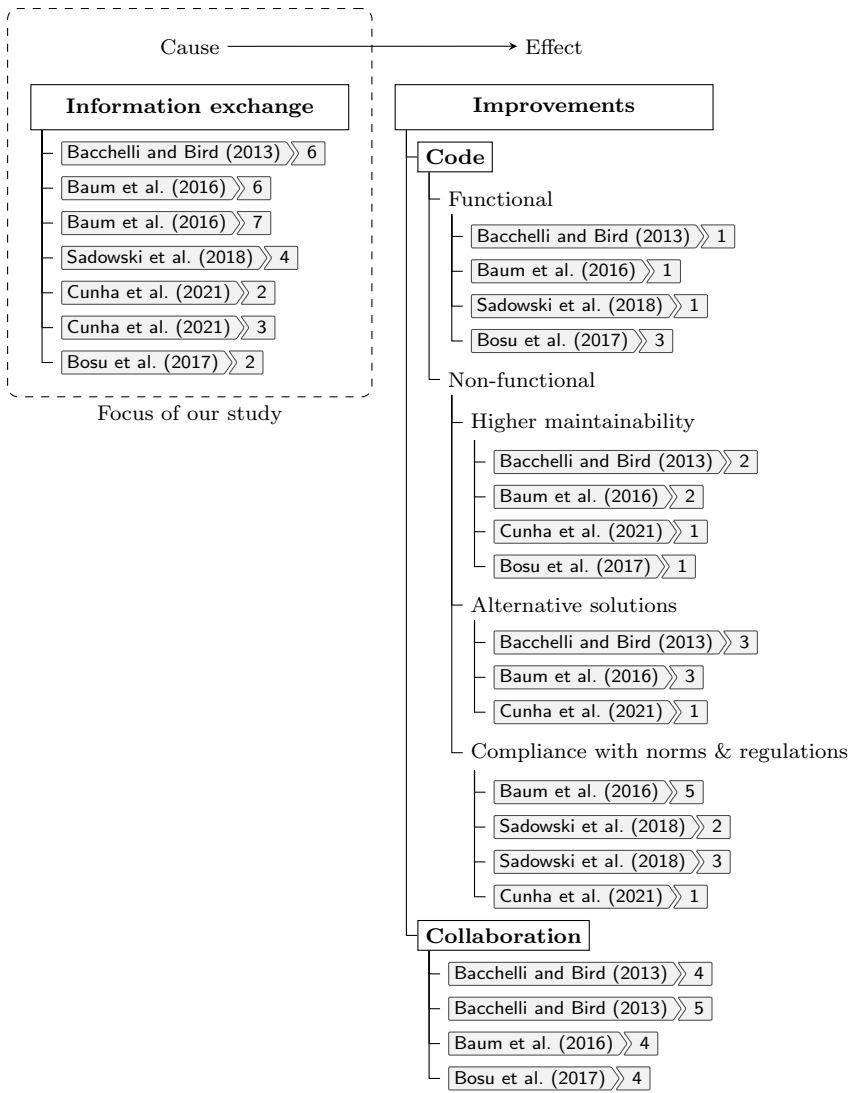

All qualitative studies reported the finding that code review is expected to exchange information. In our synthesis, we distinguish between information exchange as the root cause for the expected improvements on the one side and the expected improvements through the information exchange on the other side. We may assume that all reported and expected improvements are caused by the information exchange through code review: None of the improvements would be possible without exchanging information among the code review participants. \Cref{fig:synthesis} presents our synthesis of expectations and motivations towards code review reflecting this distinction. 

In detail, we found the expected improvements through information exchange in code review either to be related to code or to collaboration. We grouped the findings related to code into functional (identification of defects) and non-functional code improvement. The latter contains three groups of improvements: (1) alternative solutions and their discussions, (2) higher maintainability, and (3) compliance with norms and regulations. The compliance is not limited to regulations (e.g., from regulatory environments such as medical or automotive software development) but also includes organizational norms and practices directed at code. Closely related to the organizational norms and practices is the second group of expected improvements through information exchange in code review, which is collaboration. This also includes team awareness, a sense of shared code ownership, and community building.

\subsection{Measuring Information Exchange}

To the best of our knowledge (or information), there are only two cases so far to quantify knowledge sharing in code review.

The first case of measuring knowledge sharing (or information exchange) in code review was provided by \cite{Rigby2013}. The authors extended the expertise measure proposed by \cite{Mockus2002}. The study contrasts the number of files a developer has modified with the number of files the developer knows about (submitted files $\cup$ reviewed files) and found a substantial increase in the number of files a developer knows about exclusively through code review. 

The second case of measuring knowledge spreading (or information exchange) is presented by \cite{Sadowski2018}, the case study at Google discussed previously. The study reports (a) the number of comments per change a change author receives over tenure at Google and (b) the median number of files edited, reviewed, and both---as suggested by \cite{Rigby2013}. The study finds that the more senior a code change author is, the fewer code comments he or she gets. The authors ``postulate that this decrease in commenting results from reviewers needing to ask fewer questions as they build familiarity with the codebase and corroborates the hypothesis that the educational aspect of code review may pay off over time.'' In its second measurement, the study reproduces the measurements of \cite{Rigby2013} but reports it over tenure months at Google. The plot shows that reviewed and edited files are distinct sets to a large degree. 

However, we found the following limitations in the measurement applied to prior work:
\begin{itemize}
\item We are unaware of empirical evidence that exposure to files in code review would reliably lead to improved developer fluency.
\item Findings like \menu{\cite{Bacchelli2013}}{4} (team awareness and transparency) and \menu{\cite{Bacchelli2013}}{5} (share code ownership) cannot be measured.
\item The explanatory power of both measurements is limited since the authors set arbitrary boundaries: \cite{Rigby2013} excluded changes and reviews that contain more than ten files, and \cite{Sadowski2018} limited the tenure to 18 months and aggregated the tenure by three months. 
\end{itemize}

For our approach, therefore, we need to subsume all notions of knowledge by using the broader concept of information and its exchange since information encodes knowledge of all types, including meta-information, such as the social ties between developers. This subsumption was also used in \cite{Dorner2022} to validate the simulation model, showing the importance of time for measuring and analyzing information diffusion.

\subsection{Information Diffusion}

Information diffusion, the spread of information among humans, has been researched in different disciplines and for encodings of information, for example, tweets \citep{ Anger2011}, memes \citep{Leskovec2009}, blog posts \citep{Goetz2009}, or e-mail chain letters \citep{Liben-Nowell2008}. 

However, information in code review is more fine-grained and significantly harder to identify and trace than forwarded tweets, memes, blog posts, or e-mails. Therefore, \cite{Dorner2022} proposed and validated a simulation model for information diffusion tailored to code review without tracing identifiable information but focussing on the communication network emerging from code review. In detail, we modelled the code review network emerging from code review as time-varying hypergraph, where the nodes are code review participants and the (hyper)edges are code reviews that connect the participants over time. For more details, we refer the reader to \Cref{sec:simulation_model} as we reuse the simulation model from \cite{Dorner2022} in this study.



\subsection{Simulations as Empirical Research Method}

We conduct our study as an \emph{in-silico} experiment, in which we simulate an artificial information diffusion and measure the resulting traces generated by the spread of the information. Given the rarity of simulations in the empirical software engineering community, we motivate the explicit choice of that as our (empirical) research method. 

Generally speaking, an in-silico experiment is an experiment performed in a computer simulation (on silicon). In contrast to other types of experiments (see \Cref{tab:simulation_types}), all parts of the experiment, i.e., the actors, their behaviours, and the context (borrowed from \cite{Stol2018}), are modelled explicitly as computer (software) model. In-vitro experiments model the context other than using a computer model.

\begin{table}
\centering
\caption{A comparison of \emph{in-vivo}, \emph{in-vitro}, \emph{in-virtuo}, and \emph{in-silico} experiments.}
\label{tab:simulation_types}
\begin{NiceTabular}{@{}lllll@{}}[name=table]
\CodeBefore [create-cell-nodes]
\tikz{
	\node (n1) [inner sep=2pt, fit= (3-5) (5-5)] {} ;
    \node (n2) [inner sep=2pt, fit= (5-4)] {} ;
	\draw[rounded corners = 1mm, fill=yellow!50] (n1.north) -| (n1.south east) -- (n2.south west) -- (n2.north west) -- (n2.north east) -| (n1.north west) -- cycle;
	\node (n3) [draw, inner sep=2pt, rounded corners = 1mm, fit= (5-3)] {};
	\draw (n1.east) to[out=0, in=90] (8,-1.5) node[text width=2.5cm, fill=white, text centered] {as computer \\(software) model};
	\draw (n3.south) to[out=-90, in=180] (5,-1.5) node[text width=2cm, fill=white] {otherwise};
}
\Body
\toprule
& \multicolumn{4}{l}{Experiment} \\ 
\cmidrule(r){2-5}
& {in-vivo} & {in-vitro} & {in-virtuo} & {in-silico} \\
\midrule
{Actor} & natural & natural & natural & modelled \\
{Behaviour} & natural & natural & natural & modelled \\
{Context} & natural & modelled & modelled & modelled \\
\bottomrule
\\
\end{NiceTabular}
\end{table}

Those more traditional experiments in software engineering would have been too complex, too expensive, too lengthy, or simply not possible or accessible otherwise. Following \cite{Muller2008} ``Simulation models are like virtual laboratories where hypotheses about observed problems can be tested, and corrective policies can be experimented with before they are implemented in the real system.'' Those attributes match the objectives and setting of our research.

Simulation models have been applied in different research fields of software engineering, e.g., process engineering, risk management, and quality assurance \citep{Muller2008}.

The role of simulations as an empirical method is, however, still often subject to some form of prejudice but also subject to ongoing more philosophical debates. \cite{Stol2018}, for example, positioned computer simulations in their ABC framework in a non-empirical setting because, as the authors argue: ``while variables can be modelled and manipulated based on the rules that are defined within the computer simulation, the researcher does not make any new empirical observations of the behavior of outside actors in a real-world setting (whether these are human participants or systems)'' \citep{Stol2018}. Without discussing the role of simulations in the empirical software engineering community to the extent they might deserve, however, we still argue for their suitability as an evidence-based (empirical) approach in our context where observations would otherwise not be possible (or, at least, not realistic). 

We consider computer simulations as an empirical research method, the same as done in other disciplines and inter-disciplines (where, for instance, climate simulations are the first-class citizens in the set of research methods). Empirical research methods are ``research approaches to gather observations and evidence from the real world''\citep{Stol2018} and same as in other empirical research methods, in simulation models, we build the models based on real-world observations and make conclusions based on the empirical observations along the execution (in our case, of the simulations). These simulations are abstractions from the real world---same as the (often implicit) theoretical models underlying quasi-controlled (in-vitro) experiments. Simulations and their underlying models further abstract from (and make explicit) complex systems and make observations and evidence possible in situations where more traditional experiments are rendered infeasible (e.g., too expensive, dangerous, too long, or not accessible) or simply impossible at all; for instance, observations when exploring the capabilities of real-world communication networks with thousands of developers as done in the simulation study presented in the manuscript at hands.

Needless to elaborate, a certain abstraction from the real world is inherent to all empirical research methods, either in the form of explicit models or implicit assumptions. Like every measurement, the models we create come with certain accuracy and precision---with a certain quality. However, we may still argue that the quality of a research method does not necessarily decide upon whether it qualifies as empiricism or not but rather the underlying constructs and their (evidence-based) sources. To avoid surreal models and ensure the quality of a model, however, the modelling itself needs to be guided by quality assurance in verification and validation, and the sample used needs to be realistic; both would, in turn, be in tune with the underlying arguments by available positionings such as the one by \cite{Stol2018}. To increase the transparency in the quality of our simulations, we further disclose all developed software components as a replication package, also including the industrial communication networks we used as a sample.

\section{Experimental Design}\label{sec:experimental_design}

In this section, we describe the design of our \emph{in-silico} experiment that evaluates and quantifies how far (RQ~1) and how fast (RQ~2) information can diffuse in code review. 

The underlying idea of our experiment is simple: We create communication networks emerging from code reviews and then measure the minimal paths between the code review participants. The cardinality of reachable participants indicates how far (RQ~1) information can spread, and distances between participants indicate how fast (RQ~2) information can spread in code review. Since we used minimal paths and created the communication networks under best-case assumptions, the results describe the upper bound of information diffusion in code review.

Yet, since communication, and, therefore, information diffusion, is (1) inherently a time-dependent process that is (2) not necessarily bilateral---often more than two participants exchange information in a code review---, traditional graphs are not capable of rendering information diffusion without dramatically overestimating information diffusion \citep{Dorner2022}. Therefore, we use time-varying hypergraphs to model the communication network and measure the shortest paths of all vertices within those networks. Since a hypergraph is a generalization of a traditional graph, traditional graph algorithms (i.e., Dijkstra's algorithm) can be used to determine minimal-path distance. 

The connotation of minimal is two-fold in time-varying hypergraphs: A distance in time-varying hypergraphs between two vertices has not only a topological but also a temporal perspective. This allows us to measure not only the topologically minimal but also the temporal distance between vertices. Both distance types result in the same set of reachable participants, which we use to answer RQ~1. 

Since all models are abstractions and, accordingly, simplifications of reality, the quality of an \emph{in-silico} experiment highly depends on the quality of the simulation model and its parametrization. Therefore, we provide a more elaborate description of our simulation model, which was originally proposed and partially validated in \cite{Dorner2022}, in \Cref{sec:simulation_model} and its parametrization of its computer model by empirical code review data (\Cref{sec:parametrization}).

\subsection{Simulation Model}\label{sec:simulation_model}

In general, a simulation model consists of two components: (1) the conceptual model describing our derivations and assumptions and (2) the computer model as the implementation of the conceptual model. The following two subsections describe each component in detail.

\subsubsection{Conceptual Model}\label{sec:conceptual_model}

In the following, we describe how we conceptually model the communication networks from code review discussions and the information diffusion within those communication networks.

\paragraph{Communication Network}\label{sec:communication_network}

Communication, the purposeful, intentional, and active exchange of information among humans, does not happen in the void. It requires a channel to exchange information. A \emph{communication channel} is a conduit for exchanging information among communication participants. Those channels are 

\begin{enumerate}
\item {\itshape multiplexing}---A channel connects all communication participants sending and receiving information.
\item {\itshape reciprocal}---The sender of information also receives information and the receiver also sends information. The information exchange converges. This can be in the form of feedback, queries, or acknowledgments. Pure broadcasting without any form of feedback does not satisfy our definition of communication.
\item {\itshape concurrent}---Although a human can only feed into and consume from one channel at a time, multiple concurrent channels are usually used. 
\item {\itshape time-dependent}---Channels are not available all the time; after the information is transmitted, the channels are closed.
\end{enumerate}

Channels group and structure the information for the communication participants over time and content. Over time, the set of all communication channels forms a communication network among the communication participants.

In the context of our study on information diffusion, a communication channel is a discussion in a merge (or pull) request. A channel for a code review on a merge request begins with the initial submission and ends with the merge in case of an acceptance or a rejection. All participants of the review of the merge request feed information into the channel and, thereby, are connected through this channel and exchange information they communicate. After the code review is completed and the discussion has converged, the channel is closed and archived, and no new information becomes explicit and could emerge. However, a closed channel is usually not deleted but archived and is still available for passive information gathering. We do not intend to model this passive absorption of information from archived channels by retrospection with our model. For this line of research, we recommend the work by \cite{Pascarella2018} as further reading.

From the previous postulates on channel-based communication, we derive our mathematical model: Each communication medium forms an undirected, time-varying hypergraph in which hyperedges represent communication channels. Those hyperedges are available over time and make the hypergraph time-dependent. Additionally, we allow parallel hyperedges\footnote{This makes the hypergraph formally a \emph{multi-hypergraph} \cite{Ouvrard2020}. However, we consider the difference between a hypergraph and a multi-hypergraph as marginal since it is grounded in set theory. Sets do not allow multiple instances of the elements. Therefore, instead of a set of hyperedges, we use a multiset of hyperedges that allows multiple instances of the hyperedge.}---although unlikely, multiple parallel communication channels can emerge between the same participants at the same time but in different contexts.

Such an undirected, time-varying hypergraph reflects all four basic attributes of channel-based communication: 

\begin{itemize}
\item {\itshape multiplexing}---since a single hyperedge connects multiple vertices,
\item {\itshape concurrent}---since (multi-)hypergraphs allow parallel hyperedges,
\item {\itshape reciprocal}---since the hypergraph is undirected, information is exchanged in both directions, and
\item {\itshape time-dependent}---since the hypergraph is time-varying. 
\end{itemize}

In detail, we define the channel-based communication model for information diffusion in an observation window $\mathcal{T}$ to be an undirected time-varying hypergraph 
\[\mathcal{H} = (V, \mathcal{E}, \rho, \xi, \psi)\] 
where

\begin{itemize}
\item $V$ is the set of all human participants in the communication as vertices
\item $\mathcal{E}$ is a multiset (parallel edges are permitted) of all communication channels as hyperedges, 
\item $\rho$ is the \emph{hyperedge presence function} indicating whether a communication channel is active at a given time,
\item $\xi \colon E \times \mathcal{T} \rightarrow \mathbb{T}$, called \emph{latency function}, indicating the duration to exchange information among communication participants within a communication channel (hyperedge),
\item $\psi \colon V \times \mathcal{T} \rightarrow \{0, 1\}$, called \emph{vertex presence function}, indicating whether a given vertex is available at a given time. 
\end{itemize}

\paragraph{Information Diffusion}\label{sec:information_diffusion}

The time-respecting routes through the communication network are potential \emph{information diffusion}, the spread of information. To estimate the upper bound of information diffusion and, thereby, answer both of our research questions, we measure the distances between the participants under best-case assumptions. 

For information diffusion in code review, we made the following assumptions:

\begin{itemize}
\item {\itshape Channel-based}---Information can only be exchanged along the information channels that emerged from code review. The information exchange is considered to be completed when the channel is closed. 
\item {\itshape Perfect caching}---All code review participants can remember and cache all information in all code reviews they participate in within the considered time frame. 
\item {\itshape Perfect diffusion}---All participants instantly pass on information at any occasion in all available communication channels in code review. 
\item {\itshape Information diffusion only in code review}---For this simulation, we assume that information gained from discussions in code review diffuses only through code review. 
\item {\itshape Information availability}---To have a common starting point and make the results comparable, the information to be diffused in the network is already available to the participant, which is the origin of the information diffusion process. 
\end{itemize}

Our assumptions make the results of the information diffusion a best-case scenario. Although the assumptions do not likely result in actual, real-world information diffusion, they serve well the scope of our study, namely to quantify the upper bound of information diffusion. 

The possible routes through the communication network describe how information can spread through a communication network. Those routes are time-sensitive: a piece of information gained from a communication channel (i.e., a code review discussion) can be shared and exchanged in all subsequent communication channels but not in prior, closed communication channels. 

Mathematically, those routes are time-respecting walks, so-called \emph{journeys}, in a time-varying hypergraph representing the communication network. A journey is a sequence of tuples 
\[ \mathcal{J} = \left\{(e_1, t_1), (e_2, t_2), \dots, (e_k, t_k),\right\} \] 
such that $\{e_1, e_2, \dots, e_k\}$ is a walk in $\mathcal{H}$ with $\rho(e_i, t_i) = 1$ and $t_{i+1} > t_i + \xi(e_i, t_i)$ for all $i < k$.

We define $\mathcal{J}^*_{\mathcal{H}}$ the set of all possible journeys in a time-varying graph $\mathcal{H}$ and $\mathcal{J}^*_{(u, v)} \in \mathcal{J}^*_{\mathcal{H}}$ the journeys between vertices $u$ and $v$. If $\mathcal{J}^*_{(u, v)} \neq \emptyset$, $u$ can reach $v$, or in short notation $u \leadsto v$.\footnote{In general, journeys are not symmetric and transitive---regardless of whether the hypergraph is directed or undirected: $u \leadsto v \centernot\Leftrightarrow v \leadsto u$.} Given a vertex $u$, the set $\{v \in V \colon u \leadsto v \}$ is called \emph{horizon} of vertex $u$.

The notion of length of a journey in time-varying hypergraphs is two-fold: Each journey has a topological distance (measured in number of hops) and temporal distance (measured in time). This gives rise to two distinct definitions of distance in a time-varying graph $\mathcal{H}$:

\begin{itemize}
\item The \emph{topological distance} from a vertex $u$ to a vertex $v$ at time $t$ is defined by $d_{u, t}(v) = \min \{\vert \mathcal{J}(u, v) \vert_{h}\}$ where the journey length is $\vert \mathcal{J}(u, v) \vert_{h} = \vert \{e_1, e_2, \dots, e_k\} \vert$. This journey is the \emph{shortest}.
\item The \emph{temporal distance} from a vertex $u$ to a vertex $v$ at time $t$ is defined by $\hat{d}_{u, t}(v) = \min \{\psi(e_k) + \xi(e_k) - \xi(e_1)\}$.\footnote{In our case, $\psi(e_k)$ is always $0$.} This journey is the \emph{fastest}.\footnote{For the interested reader, we would like to add that if the temporal distance is not defined for a relative time but for an absolute time $\hat{d}_{u, t}(v) = \min \{\psi(e_k) + \xi(e_k)\}$, the journey is called \emph{foremost}. For this line of research, the foremost journeys are not used.}
\end{itemize}

With this conceptual model and its mathematical background, we are now able to answer both research questions by measuring two characteristics of all possible routes through the communication network:
\begin{itemize}
\item The distribution of the horizon of each participant in a communication network represents how far information can spread (RQ~1). 
\item The distribution of all shortest and fastest journeys between all participants in a communication network answers how fast information can spread in code review (RQ~2). We measure how fast information can spread in code review in terms of the topological distance (minimal number of code reviews required to spread information between two code review participants) and the temporal distance (minimal timespan to spread information between two code review participants).
\end{itemize}

Those measurements within code review communication networks will result in the upper bound of information diffusion in code review.

\subsubsection{Computer Model}\label{sec:computer_model}

Since our mathematical model is not trivial and lacks performant tool support for time-varying hypergraphs, we dedicate this section to the computer model and the implementation of the mathematical model described previously.

Time-varying hypergraphs are a novel concept; therefore, we cannot rely on existing toolings. We implemented the time-hypergraph as an equivalent bipartite graph: The hypergraph vertices and hyperedges become two sets of vertices of the bipartite graph. The vertices of those disjoint sets are connected if a hypergraph edge is part of the hyperedge. \Cref{fig:intro} shows a graphical description of the equivalence of hypergraphs and bipartite graphs.

\begin{figure*} 
\centering
\begin{subfigure}[t]{0.48\textwidth}
\captionsetup{skip=1em}%
\centering
\begin{tikzpicture}
\tikzstyle{hedge}=[draw, circle, minimum size=8mm, fill=white];

\begin{scope}[rotate=-5, xshift=0.0cm]
\node at (1,1) (v1) {};
\node at (2,2) (v2) {};
\node at (2,0) (v3) {};

\end{scope}

\begin{scope}[rotate=5, yshift=0.0cm]
\node at (4,2) (v4) {};
\node at (4,-1) (v5) {};
\node at (5,0.5) (v6) {};
\end{scope}

\node at (0,-2) {}; 

\draw[draw, fill=e1_color, fill opacity=0.5] \hedgem{v1}{v2}{v3}{6mm};
\draw[draw, fill=e2_color, fill opacity=0.5] \hedgeii{v2}{v4}{6mm};
\draw[draw, fill=e4_color, fill opacity=0.5] \hedgem{v6}{v5}{v4}{6mm};
\draw[draw, fill=e3_color, fill opacity=0.5] \hedgem{v6}{v5}{v3}{6mm};

\node[hedge] at (v1) {$v_1$};
\node[hedge] at (v2) {$v_2$};
\node[hedge] at (v3) {$v_3$};
\node[hedge] at (v4) {$v_4$};
\node[hedge] at (v5) {$v_5$};
\node[hedge] at (v6) {$v_6$};

\node at (barycentric cs:v1=1,v2=1,v3=1) {$e_1$};
\node at (barycentric cs:v2=1,v4=1) {$e_2$};
\node at (barycentric cs:v3=1,v5=1) {$e_3$};
\node at (barycentric cs:v4=1,v6=1) {$e_4$};

\end{tikzpicture} 
\caption{An example time-varying hypergraph whose so-called hyperedges (denoted by $e_\square$) can link any arbitrary number of vertices (denoted by $v_\square$): For example, hyperedge $e_3$ connects three vertices. The horizon and minimal paths of vertex depend highly on the temporal order of the hyperedges: for example, the horizon of $v_1$ contains all vertices if the temporal availabilities of the hyperedges are $e_1 < e_2 < e_4 < e_3$, but none if $e_1 > e_2 \geq e_3$.}
\label{fig:example_hypergraph}
\end{subfigure}%
\hfill%
\begin{subfigure}[t]{0.48\textwidth}
\captionsetup{skip=1em}%
\centering
\begin{tikzpicture}
\tikzstyle{hedge}=[draw, circle, minimum size=8mm, fill=white];

\begin{scope}[local bounding box=hypergraph, anchor=center, start chain=going below, node distance=2mm]
\foreach \i in {1,2,...,6}
  \node[on chain, hedge] (v\i) {$v_\i$}; 
\end{scope}

\begin{scope}[draw, anchor=center, shift={(2,-0.1)}, start chain=going below, node distance=8mm]
\foreach \e/\c in {1/e1_color, 2/e2_color, 3/e3_color, 4/e4_color}
  \node[on chain, draw, circle, fill=\c!50, minimum size=8mm] (e\e) {\textcolor{black}{$e_\e$}};
\end{scope}

\draw[e1_color] (v1) -- (e1);
\draw[e1_color] (v2) -- (e1);
\draw[e1_color] (v3) -- (e1);

\draw[e2_color] (v2) -- (e2);
\draw[e2_color] (v4) -- (e2);

\draw[e3_color] (v3) -- (e3);
\draw[e3_color] (v5) -- (e3);
\draw[e3_color] (v6) -- (e3);

\draw[e4_color] (v4) -- (e4);
\draw[e4_color] (v5) -- (e4);
\draw[e4_color] (v6) -- (e4);

\draw [decorate, decoration={brace, mirror, amplitude=5pt, raise=4ex}] (v1.north west) -- (v6.south west) node[midway, xshift=-10mm]{\rotatebox{90}{Vertices}};

\draw [decorate, decoration={brace, amplitude=5pt, raise=4ex}] (e1.north east) -- (e4.south east) node[midway, xshift=10mm]{\rotatebox{90}{Hyperedges}};



\end{tikzpicture}
\caption{Any hypergraph can be transformed into an equivalent bipartite graph: The hyperedges and the vertices from the time-varying hypergraph from (a) become the two distinct sets of vertices of a bipartite graph.}
\label{fig:bipartite_graph}
\end{subfigure}

\caption{An example hypergraph (a) and its bipartite-graph equivalent (b).}
\label{fig:intro}
\end{figure*} 

We use a modified Dijkstra's algorithm to find the minimal journeys for each vertex (participant) in the time-varying hypergraph. Dijkstra's algorithm is asymptotically the fastest known single-source shortest-path algorithm for arbitrary directed graphs with unbounded non-negative weights. In contrast to its original form, our implementation finds both the shortest (a topological distance) and fastest (a temporal distance) journeys in time-varying hypergraphs.\footnote{For future applications, our implementation of Dijkstra's algorithm can also find any foremost journey.}

Since Dijkstra's algorithm can be seen as a generalization of a breadth-first search for unweighted graphs, we can identify not only the minimal paths but also the horizon of each participant in the communication network in one computation. 

The algorithm is integrated into our computer model and implemented in Python. For more implementation details and performance considerations, we refer the reader to our replication package, including its documentation \citep{Dorner2023software,Dorner2023data}. Because both time-varying hypergraphs as the data model and the extended Dijkstra's algorithm are novel, we ensure the computational model accurately represents the underlying mathematical model and the correctness of our Dijkstra implementation and its results through the following quality assurance measures: 

\begin{itemize}
\item \textit{Code walk-throughs}---We independently conducted code walk-throughs through the simulation code with three Python and graph experts. 
\item \textit{Comprehensive test setup}---The simulation code has a test coverage of over \qty{99}{\percent}.
\item \textit{Code readability and documentation}---We provide comprehensive documentation on the usage and design decisions to enable broad use and further development. We followed the standard Python programming style guidelines PEP8 to ensure consistency and readability. 
\item \textit{Publicly available and open source}---The model parameterization and simulation code \cite{Dorner2023software} as well as the results \cite{Dorner2023data} for replications and reproductions are publicly available.
\end{itemize}

\subsection{Model Parametrization}\label{sec:parametrization}

Instead of a theoretical or probabilistic parametrization, we parametrize our simulation model with empirical code review systems from three industrial partners: Microsoft, Spotify, and Trivago. 

In the following, we describe our sampling strategy for selecting suitable code review systems in industry and the code review data extraction for parametrizing our simulation model.

\subsubsection{Sampling}\label{sec:sampling}

Communication networks do not emerge in the void. They form around software development. As motivated in the introduction, we focus in our study on industrial software development since we believe that results found in open-source projects cannot be directly transferred due to the differences in governance structures, incentives, and economic mechanics. Also, previous qualitative work, which we aim to complement with our work (see \Cref{sec:expectations}), considers industrial software development only.

We use a \emph{maximum variation sampling} to select suitable code review communication networks in an industrial context. A maximum (or maximum heterogeneity) variation sampling is a non-probabilistic, purposive sampling technique that chooses a sample to maximize the range of perspectives investigated in the study in order to identify important shared patterns that cut across cases and derive their significance from having emerged out of heterogeneity \citep{Teddlie2009}. 

We aim for representativeness on two specific dimensions: 

\begin{itemize}
\item \textit{Code review system size}---To avoid a bias introduced by network effects, we required communication networks emerging from different sizes of code review. The size of a communication network can be measured in terms of the number (hyper)edges (corresponding to the number of code reviews) or vertices (corresponding to the number of participants). In our sample, we use a small (Trivago), mid-sized (Spotify), and large (Microsoft) code review system (see \Cref{tab:sample}). The size classification in small, mid-sized, and large code review systems is arguably arbitrary and relative to the code review systems in our sample rather than following a general norm that, to the best of our knowledge, does not exist.
\item \textit{Code review tool}---In particular, since \cite{Baum20161} suggested code review tool in use as a main factor shaping code review in industry, we aim to minimize the code review tool bias for the results and require our sample to contain a diverse set of code review tools. Our sample contains three different code review tools: BitBucket, GitHub, and CodeFlow. 
\end{itemize}

In alignment with the qualitative prior work, we explicitly excluded the different manifestations in code review practices as a sampling dimension. To ensure that the results are comparable within and among the selected contexts and to ease the data extraction, we restrict our population to having a single, central code review tool in use. This means our population is any industrial software development company with a single, centralized code review tool. Like any purposive sampling technique, the maximum variation sampling does not require a sampling frame \citep{Baltes2022}. 

From this population, we drew a sample of three industrial cases: Microsoft, Spotify, and Trivago. \Cref{tab:sample} provides an overview of our sample of code review systems and the dimension of representativeness. We describe the cases in our sample in more detail in the following subsections.

\begin{table}
\centering
\caption{Our sample of code review systems with respect to the two dimensions of representativeness: code review system size and tooling.}
\label{tab:sample}
\begin{tabular}{lllll}
\toprule
{} & \multicolumn{3}{l}{Code review system size} & \multirow{2}{*}{Tooling}\\ \cmidrule{2-4}
{} & Classification & Code reviews & Participants & {}\\
\midrule
Trivago & small & \num{2442} & \num{364} & BitBucket \\
Spotify & mid-sized & \num{22504} & \num{1730} & GitHub \\
Microsoft & large & \num{309740} & \num{37103} & CodeFlow \\
\bottomrule
\end{tabular}
\end{table}

\paragraph{Microsoft} 
Microsoft is a multinational enterprise that produces computer software, consumer electronics, personal computers, and related services and is based in the USA. We extracted the data from Microsoft's internal code review tool \emph{CodeFlow} \citep{Bosu2015} run by Azure DevOps service. Although not Microsoft's only code review tool, it covers the vast majority of the company's code review activity. The dataset contains \num{37103} code review participants and \num{309740} code reviews within the observation window between 2020-02-03 and 2020-03-02. 

\paragraph{Spotify} 
Spotify is a multinational enterprise based in Sweden that develops a multimedia streaming platform. We extracted all internal pull requests and their related comments within the observation window between 2020-02-03 and 2020-03-02 from Spotify's GitHub Enterprise instance, the central tool for software development at Spotify. The dataset contains \num{1730} code review participants and \num{22504} code reviews. 

\paragraph{Trivago}
Trivago is a German company developing an accommodation search engine. As a code review tool, Trivago used Bitbucket during the observation window between 2019-11-04 and 2019-12-01. The dataset contains \num{364} code review participants and \num{2442} code reviews.

\subsubsection{Data Collection}\label{sec:data_collection}

We extract all human interactions with the code review discussions within four consecutive calendar weeks from the single, central code review tools in each industrial context. 

We define a code review interaction as any non-trivial contribution to the code review discussion: Creating, editing, approving or closing, and commenting on a code review. For this study, we do not consider other (tool-specific) types of discussion contributions, for example, emojis or likes to a contribution to a code review. 

The beginning and end of those four-week timeframes differ and are arbitrary, but share the common attributes: All timeframes 

\begin{itemize}
\item start on a Monday and end on a Sunday,
\item have no significant discontinuities by public holidays such as Christmas, 
\item are pre-pandemic to avoid introduced noise from introduced work-from-home policies, pandemic-related restrictions, or interferences in the software development.
\end{itemize}

\Cref{tab:timeframes} lists the timeframes (each four weeks) and when the data was collected. 

\begin{table}
\centering
\begin{tabular}{lll}
\toprule
 & Observation window (four weeks) & Collected during \\
\midrule
Microsoft & 2020-02-03 to 2020-03-02 & May 2020\\
Spotify & 2020-02-03 to 2020-03-02 & March 2023\\
Trivago & 2019-11-04 to 2019-12-01 & December 2022\\
\bottomrule
\end{tabular}	
\caption{Observation window and the data collection timeframe among our cases. }
\label{tab:timeframes}
\end{table}

All non-human code-review participants and interactions (i.e., bots or automated tasks contributing to the code-review discussions) are excluded. We strictly anonymized all participants and removed all identifiable personal information to protect the privacy of all individuals. 

All data and results are publicly available \citep{Dorner2023data}.

\section{Results}\label{sec:results}

This section presents the results of our simulation as described in \Cref{sec:experimental_design} and is structured around our two research questions. 

Both research questions cover different perspectives on code review as a communication network: In RQ~1, we use a vertex-centric perspective by measuring the reachable participants (vertices) for each participant (vertex). For RQ~2, we use a hyperedge-centric perspective by measuring the topological and temporal lengths of paths through the communication network that emerges from code review.

\subsection{How far can information diffuse through code review (RQ~1)?}

As described in \Cref{sec:conceptual_model}, we answer RQ~1 by measuring the number of reachable participants for each participant in the communication network that emerges from code review. The number of reachable participants is the cardinality of each participant's horizon. In the following, we call the number of reachable participants \emph{information diffusion range}. 

To make the different code review system sizes comparable, we normalize the information diffusion range to the number of code review participants in a code review system. Mathematically, we define the normalized information diffusion range for all code review participants $u \in V$ by
 
\[
	\frac{\lvert \{v \in V \colon u \leadsto v \} \rvert}{|V|}.
\]

\begin{figure*}
\centering
\begin{tikzpicture}
\begin{axis} [ %
	ecdf axis,
	xlabel={Normalized information diffusion range after four weeks},
	ylabel={Cumulative distribution},
	xmin=0.0,
	xmax=1.0,
	xtick = {0, 0.1, 0.2, 0.3, 0.4, 0.5, 0.6, 0.7, 0.8, 0.9, 1.0},
	legend pos=north west,
]
\addplot[trivago_color, name path=trivagocov] table [x=index, y=Trivago, col sep=comma] {csv/ranges.csv} coordinate (maxtrivago);
\addlegendentry{Trivago}
\addplot[spotify_color, name path=spotifycov] table [x=index, y=Spotify, col sep=comma] {csv/ranges.csv} coordinate (maxspotify);
\addlegendentry{Spotify}
\addplot[microsoft_color, name path=microsoftcov] table [x=index, y=Microsoft, col sep=comma] {csv/ranges.csv} coordinate (maxmicrosoft);
\addlegendentry{Microsoft}

\coordinate (o) at (0,0);

\path [name path=pmin] (0, 0) -- (1, 0);
\path [name path=pmax] (0, 1) -- (1, 1);
\path [name path=p90] (0, 0.9) -- (1, 0.9);
\path [name path=p70] (0, 0.70) -- (1, 0.70);
\path [name path=p50] (0, 0.5) -- (1, 0.5);
\path [name path=p30] (0, 0.3) -- (1, 0.3);
\path [name path=p10] (0, 0.1) -- (1, 0.1);


\draw[trivago_color, thick, dashed] (maxtrivago) -- (o-|maxtrivago);
\draw[microsoft_color, thick, dashed] (maxmicrosoft) -- (o-|maxmicrosoft);

\pgfplotsinvokeforeach{30, 50, 70, 90} {
	\path[name intersections={of=trivagocov and p#1, by=ip#1}] ;
	\draw[black, |-|] (ip#1) -- node[midway] (m#1) {} (ip#1-|maxtrivago);
}

\node[right, anchor=west, fill=white] at (ip30-|maxtrivago) {$[0.67, 0.85]$};
\node[right, anchor=west, fill=white] at (ip50-|maxtrivago) {$[0.72, 0.85]$};
\node[right, anchor=west, fill=white] at (ip70-|maxtrivago) {$[0.81, 0.85]$};
\node[right, anchor=west, fill=white] at (ip90-|maxtrivago) {$[0.83, 0.85]$};
\end{axis}
\end{tikzpicture}
\caption{Cumulative distribution of information diffusion range per participant. The smallest $y$ value for a given $x$ among all three distributions indicates the upper bound of information diffusion with respect to how many participants can be reached (RQ~1). For example, \qty{30}{\percent} of all participants at Trivago can reach between \qty{0}{\percent} and \qty{67}{\percent} and \qty{70}{\percent} of the participants reach between \qty{67}{\percent} and \qty{85}{\percent} of all participants. }
\label{fig:normalized_ranges}
\end{figure*}
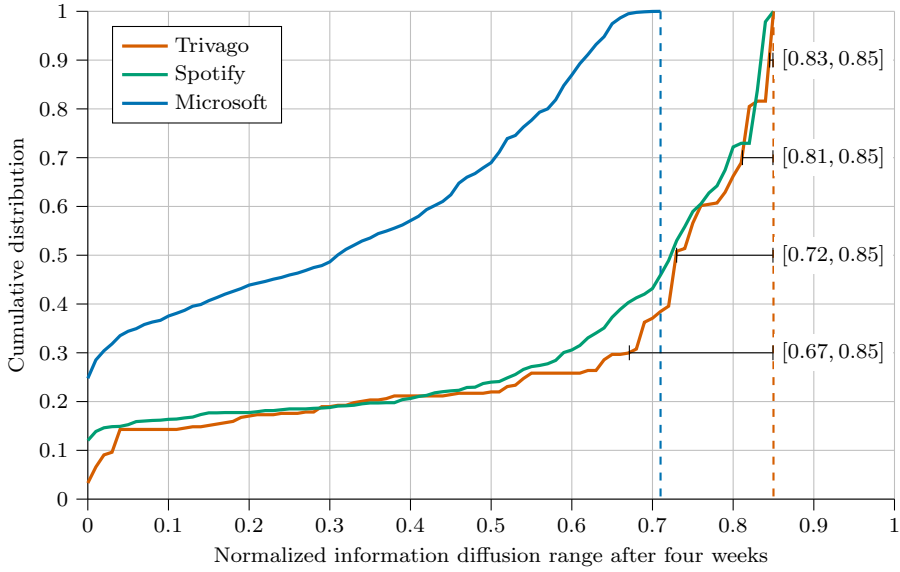

\Cref{fig:normalized_ranges} plots the empirical cumulative distribution functions (ECDF) visualizing the distributions of the information diffusion range per participant after four weeks each resulting from our simulation. 

We found the upper bound of the relative information diffusion range at Trivago's code review system. In detail, a code review participants at Trivago can reach \qty{85}{\percent} of its network at maximum. \qty{30}{\percent} of the code review participants can reach between \qty{81}{\percent} and \qty{85}{\percent}, while an average (median) code review participant can reach between \qty{72}{\percent} and \qty{85}{\percent} of all participants within the network. The code review system at Spotify generates an almost identical distribution of reachable participants. \Cref{tab:normalized_ranges} lists the ranges of horizons possible for the $p$-percentiles \numlist{0.7;0.5;0.3;0.1}. 

\begin{table*}
\centering
\caption{Cumulative distribution of normalized information diffusion range per participant for the percentiles \numlist{0.7;0.5;0.3;0.1}, and the maximum for each code review system.}
\label{tab:normalized_ranges}
\input{tex/normalized_ranges.tex}
\end{table*}

\begin{result}{result:normalized_ranges}
The code review networks at Trivago and Spotify describe almost identically the upper bound of the normalized information diffusion range: Half of the participants at Trivago and Spotify can reach between \qty{72}{\percent} and \qty{85}{\percent} of all participants within four weeks under best-case assumptions. 
\link{fig:normalized_ranges}
\end{result}

If we consider the absolute information diffusion range for each code instead review participant $u \in V$ defined by

\[\lvert \{v \in V \colon u \leadsto v \} \rvert, \]

code review participants at Microsoft's code review system can reach by far the most participants. Although the relative information diffusion range at Microsoft's code review system is significantly smaller, Microsoft's code review system sets the upper bound for the absolute information diffusion range. In detail, the code review system at Microsoft can spread information up to \num{26216} participants (\qty{71}{\percent} of the total network size), half of the code review participants can reach \num{11645} or more other participants. \Cref{tab:range} lists the ranges of the top percentiles. 

\begin{table*} 
\centering
\caption{Cumulative distribution of information diffusion range per participant for the percentiles \numlist{0.7;0.5;0.3;0.1}, and the maximum for each code review system.}
\label{tab:range}
\input{tex/ranges.tex}
\end{table*}

\begin{result}{result:range}
The code review network at Microsoft describes the upper bound of the absolute information diffusion range: Half of the participants at Microsoft can reach between \num{11645} and \num{26216} participants within four weeks under best-case assumptions.
\link{tab:range}
\end{result}

\subsection{How fast can information diffuse through code review (RQ~2)?}

As described in \Cref{sec:conceptual_model}, we answer RQ~2 by measuring the distances between the code review participants. We recall that the notion of distance in time-varying hypergraphs is two-fold: Each time-respecting path (journey) has a topological distance (the minimal number of hops of all journeys) and temporal distance (measured in time of all journeys). 

Therefore, we align the answers to RQ~2 with those two types of distances.

\subsubsection{Topological Distances in Code Review}

\begin{figure*}
\centering
\begin{tikzpicture}
\begin{axis}[
	ecdf axis,
	xmin=0,
	xmax=40,
	legend pos=north east,
	xlabel={Topological distance},
	ylabel={Cumulative distribution},
]
\addplot[trivago_color, name path=trivagoshortest] table [x=index, y=Trivago, col sep=comma] {csv/topological_distances.csv} coordinate (maxtrivago);
\addlegendentry{Trivago}
\addplot[spotify_color, name path=spotifyshortest] table [x=index, y=Spotify, col sep=comma] {csv/topological_distances.csv} coordinate (maxspotify);
\addlegendentry{Spotify}
\addplot[microsoft_color, name path=microsoftshortest] table [x=index, y=Microsoft, col sep=comma] {csv/topological_distances.csv} coordinate (maxmicrosoft);
\addlegendentry{Microsoft}

\coordinate (o) at (0,0);

\draw[trivago_color, thick, dashed] (maxtrivago) -- (o-|maxtrivago);
\draw[spotify_color, thick, dashed] (maxspotify) -- (o-|maxspotify);
\draw[microsoft_color, thick, dashed] (maxmicrosoft) -- (o-|maxmicrosoft);

\end{axis}
\end{tikzpicture}
\caption{The cumulative distribution of topological distances between participants in code review systems. The topological distance is the minimal number of code reviews (hops) required to spread information from one code review participant to another. }
\label{fig:topological_distances}
\end{figure*}
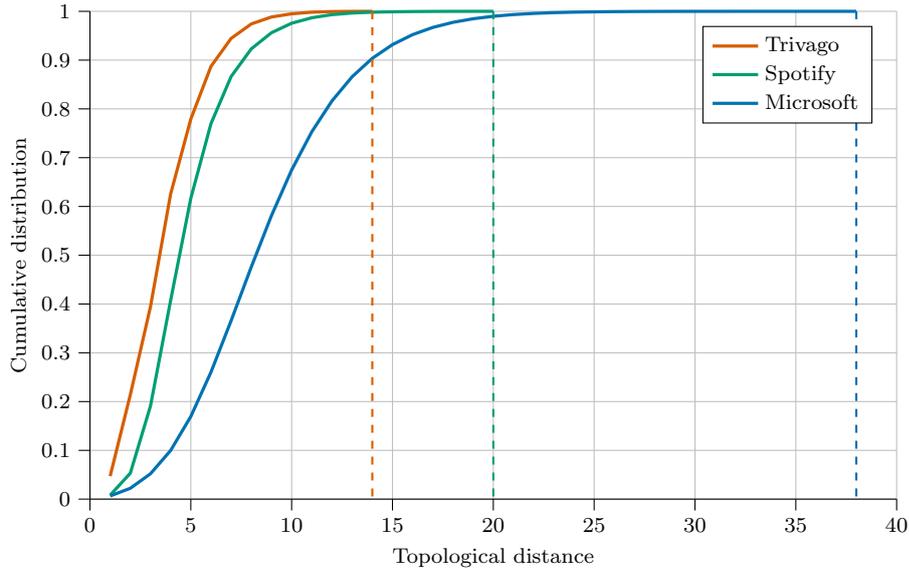

\Cref{fig:topological_distances} depicts the cumulative distribution of topological distances between code review participants among the sampled cases. 

The code review system at Trivago contains the most distances among our three cases. The average (median) distance between two participants at Trivago is three, at Spotify four, and at Microsoft eight hops. \qty{60}{\percent} of all distances at Trivago and Spotify are shorter or equal to five code reviews. The maximum distance per case is 14 for Trivago, 20 for Spotify, and 38 for Microsoft. 

\begin{result}{result:topological_distances}
Trivago's code review system describes the upper bound on how fast information can spread through code review: About \qty{75}{\percent} of all distances in Trivago's code review system between code review participants are shorter than five code reviews. 
\link{fig:topological_distances}
\end{result}

\subsubsection{Temporal Distances in Code Review}

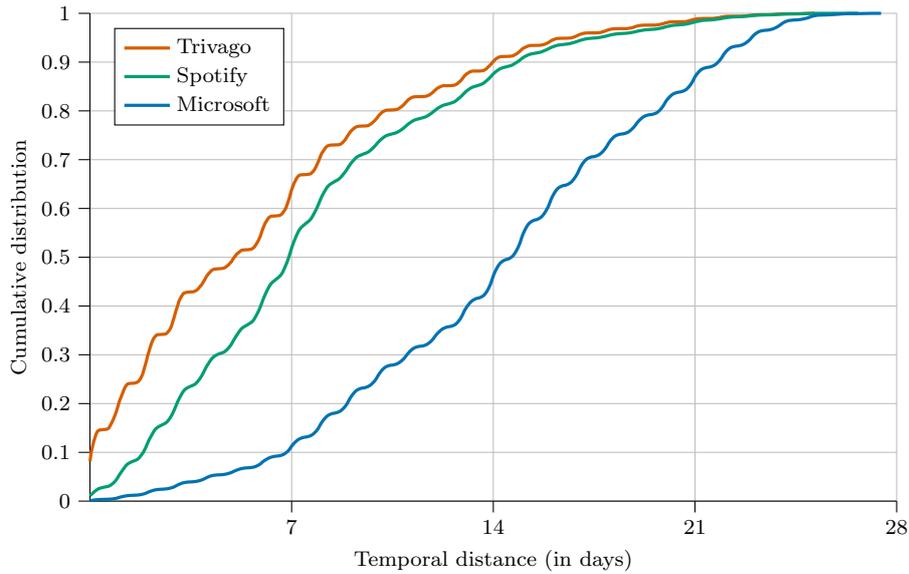
\begin{figure*}\centering
\begin{tikzpicture}
\begin{axis}[
	ecdf axis,
	xmax=672,
	xtick={168, 336, 504, 672},
	xticklabels={7, 14, 21, 28},
	legend pos=north west,
	xlabel={Temporal distance (in days)},
	ylabel={Cumulative distribution},
]
\addplot[trivago_color, name path=trivagocov] table [x=index, y=Trivago, col sep=comma] {csv/temporal_distances.csv};
\addlegendentry{Trivago}
\addplot[spotify_color, name path global=spotifycov] table [x=index, y=Spotify, col sep=comma] {csv/temporal_distances.csv};
\addlegendentry{Spotify}
\addplot[microsoft_color, name path=microsoftcov] table [x=index, y=Microsoft, col sep=comma] {csv/temporal_distances.csv};
\addlegendentry{Microsoft}
\end{axis}
\end{tikzpicture}
\caption{The cumulative distribution of minimal temporal distances between participants in code review systems. The temporal distance is the minimal duration required to spread information from one participant to another.}
\label{fig:temporal_distances}
\end{figure*}

The other type of distance in time-varying hypergraphs is the temporal distance. The fastest time-varying path is the path between two code review participants with the minimal (relative) temporal distance between two code review participants describes the minimal timespan to spread information from one participant to another. Due to our observation window, the temporal distance cannot exceed four weeks in our measurement. \Cref{fig:temporal_distances} depicts the cumulative distribution of the relative temporal distances between the code review participants in our sample.

The average (median) temporal distance between two code review participants at Trivago or Spotify is less than seven days, while a code review participant a Microsoft takes more than 14 days, which is still in the observation window of four weeks. 

\begin{result}{result:temporal_distances}
Trivago's code review system describes the upper bound on how fast information can spread through code review concerning the relative temporal distance: The average (median) temporal distance between two code review participants at Trivago are five days.
\link{fig:temporal_distances}
\end{result}

\section{Discussion}\label{sec:discussion}

Both research questions cover two different and complementary perspectives on communication networks that emerge from code review. RQ~1 captures a vertex-centric perspective on code review focusing on the participants as nodes in the communication network and their horizon. RQ~2 captures a (hyper)edge-centric perspective focussing on the length of minimal paths through the networks, representing the code reviews as communication channels connecting the participants. Also, for this section, we group our discussion around the two research questions.

\subsection{How far can information diffuse through code review (RQ~1)?}

We found a relative upper bound on how far information can spread through code review for small and mid-sized code review systems and an absolute upper bound for large code review systems (see \Cref{result:normalized_ranges,result:range}). 

Although the code review system at Trivago describes the upper bound on how far information can spread through code review (RQ~1), the code review system at Spotify has almost identical distributions of normalized information diffusion range---despite a different tooling and different total network size. Both code review systems define the upper bound of how far information can spread relative to the network size (see \Cref{result:normalized_ranges}). We consider that as a first indicator that the choice of tooling is secondary. However, although the similarity between the small and mid-sized code review systems is striking, this study was not designed to examine patterns among the code review systems. Thus, we cannot exclude a random correlation. 

Microsoft's code review system, as the largest code review system, however, describes the absolute upper bound on how far information can spread (see \Cref{result:range}). 

We are surprised by those two results as we expected a significantly smaller relative and absolute upper bound that is more guided by the organizational or software architectural boundaries. Although neither organizational nor software architectural information is available for this study, we assume an information diffusion beyond organizational or software architectural boundaries among all cases because all information diffusion ranges are magnitudes larger than reasonable team sizes. Although this study does not evaluate the expected improvements of code review but focuses on the underlying expected cause of information exchange through code review (see \Cref{fig:synthesis}), this finding corroborates the expectation \menu{\cite{Bacchelli2013}}{4} towards team awareness and transparency: Information can leave the organizational boundaries in code review leading to improved collaboration.

\subsection{How fast can information diffuse through code review (RQ~2)?}

Although we found both the topological and temporal distances to be minimal in Trivago's code review system (see \Cref{result:topological_distances,result:temporal_distances}), the temporal distances we measured at Spotify and Microsoft are remarkable given their code review sizes on almost a logarithmic scale: The average (median) distance at Spotify's code review system is shorter than five code reviews (topological distance) and shorter than seven days (temporal distance); The average (median) distance at Microsoft's code review system is shorter than seven code reviews (topological distance) and shorter than about 14 days (temporal distance).

The step-like characteristics among all cases, but more prominent in Trivago's and Microsoft's code review system, indicate a common day-night and workday rhythm of the two participants connected by the fastest time-respecting path. Although the developers' locations are not available to us and the investigation is out of the scope of our study, we speculate that information diffusion in the code review systems at Trivago and Microsoft stays mostly in the same timezone. However, Spotify's code review system describes a less distinct pattern with less clear steps. Therefore, we assume an information diffusion beyond timezones at Spotify's code review system.

\section{Limitations}\label{sec:limitations}

In outlining the limitations of our study, it is crucial to acknowledge potential threats to validity. This section highlights key constraints, both internal and external, providing context for the interpretation of our findings and suggesting avenues for future research.

\subsection{Best-Case Assumptions}

As already discussed in \Cref{sec:conceptual_model}, our assumptions regarding the information diffusion make the results a best-case scenario that is unlikely to be achieved in reality: Information is unlikely to spread on every occasion or to all code review participants. Information diffusion depends on the capability of human participants to buffer, filter, and consolidate information from their minds. Since we are unaware of any prior work on those capabilities, the results remain a theoretical upper bound of information diffusion but are no information diffusion in code review.

\subsection{Non-human Code Review Activities}

Although we excluded all code review bots from the network, the effects of bot activities on the communication network still remain: 

First, we found evidence that participants (at least partially) automated the code review handling. Bots disguised as human participants can distort the results since those bots connect more code reviews and, therefore, people than humans do. After removing all known and explicitly labeled bots, we found 14 accounts that contributed to more than 500 code reviews during our observation window of four weeks. All of those were in Microsoft's code review systems. Assuming 20 workdays within our observation windows and 8 hours a day, 500 code reviews within four weeks means about three code reviews per hour on average (mean). The maximum number of contributed code reviews is \num{8249} which then corresponds to about \num{50} code reviews per hour on average (mean). We consider that code review load is possible but highly unlikely. We did not remove the questionable accounts for the following reasons:

\begin{itemize}
\item We are unaware of empirical studies reporting the upper bound of a code review load in industry. Existing prior work on workload-aware code review participant selection does not report a distribution of code review involvement, normalize the code review size to the number of involved files, or is based on open source projects \citep{AlZubaidi2020, Armstrong2017,Chouchen2021,Strand2020}. Any threshold would be, therefore, arbitrary.
\item We cannot distinguish between bot activity (for example, a one-time cleaning script of the code review) and an actual human within such a questionable account. 
\end{itemize}

An in-depth inspection was not possible as it would require a complete de-anonymization of the accounts and analysis of the content of the code reviews of those which is not covered by the study's non-disclosure agreement.

Second, bots can provide assistance for or enforce code review guidelines by selecting and informing a set of code review participants. Those bot activities shape the network drastically and are not covered by our work.

In the foreseeable future, LLM-based bots may become code review participants. On the one hand, they produce code that is required to be reviewed by a human as long as machines cannot be held liable and accountable and they can provide feedback and share important contextual information. However, these increased bot activities may increase the workload of the human reviewers and even slow down the communication through code review \citep{Wessel2021}. The promising work by \cite{Roeseler2023} can lay the foundation for understanding the impact of bots on communication, coordination, and collaboration. 

Thus, excluding those code review activities would not reflect the information diffusion in code review anymore in the near future. However, since all the observation windows for all code review systems were located before the rise of LLMs, we are convinced that excluding bot accounts is appropriate. Future work is needed to investigate the impact of non-human code review from a communication network perspective.

\subsection{Observation Window}

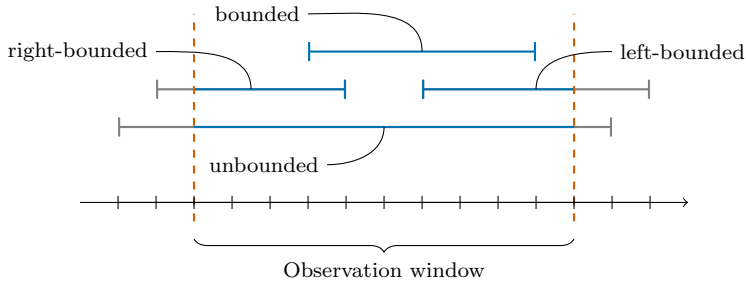
\begin{figure}
\centering
\tikzstyle{cr}=[draw, |-|, thick, gray]
\begin{tikzpicture}[x=1cm, y=0.5cm]	

\draw[cr, blue] (4.5,3) -- (7.5,3) node[midway] (bounded) {};

\draw[cr] (2.5, 2) -- (5, 2) node[midway] (right_bounded) {}; 
\draw[draw, thick, -|, blue] (3, 2) -- (5, 2); 

\draw[cr] (6,2) -- (9,2) node[midway] (left_bounded) {}; 
\draw[draw, thick, |-, blue] (6, 2) -- (8, 2); 

\draw[cr] (2, 1) -- (8.5, 1);	
\draw[draw, thick, blue] (3, 1) -- (8, 1) node[midway] (unbounded) {};

\draw[dashed, red, thick] (3, -1.5) -- (3, 4);
\draw[dashed, red, thick] (8, -1.5) -- (8, 4);
\draw [decorate,decoration={brace, amplitude=5pt, mirror, raise=4ex}] (3, -1) -- (8,-1) node[midway, yshift=-3em]{Observation window};
\draw[->] (1.5, -1) -- (9.5, -1);
\draw [
postaction={
    draw,
    decoration=ticks,
    segment length=0.5cm,
    decorate,
}
 ] (2,-1) -- (9,-1);

\draw (left_bounded.center) to[out=90, in=-180] (8.5, 3) node[right] {left-bounded};
\draw (right_bounded.center) to[out=90, in=0] (2.5, 3) node[left] {right-bounded};
\draw (unbounded.center) to[out=-90, in=0] (4.75, 0) node[left] {unbounded};
\draw (bounded.center) to[out=90, in=0] (4.5, 4) node[left] {bounded};

\end{tikzpicture}
\caption{The impact of observation windows on data completeness: The concurrent code reviews as communication channels may have started before or ended after the observed time window. Due to the cut, the communication channels may cut at their start (right-bound) or at their end (left-bound), or the channel is completely contained (bound) or not contained (unbound) in the observation window. }
\label{fig:observation_window}
\end{figure}

As for any continuous, real-world process, we only can make assumptions about windowed observations of that phenomenon. At the border of our observation windows, we have to live with some blur and uncertainty: The communication channel may have started before or ended after our observed time window. \Cref{fig:observation_window} demonstrates the problem of the observation window for an ongoing system. A channel is either
\begin{itemize}
\item unbounded (observation window does not include start or end of the channel)
\item bounded (observation window contains start and end of the channel)
\item left-bounded (observation window contains start but no end of the channel)
\item right-bounded (observation window contains end but not start of the channel)
\end{itemize}

\begin{figure*}
\centering
\begin{tikzpicture}
\begin{axis}[
height=6cm,
width=6cm,
ybar stacked,
ymin=0,
ymax=1,
enlarge x limits=0.2,
enlarge y limits=0.,
xtick=data, 
bar width=0.5cm,
xticklabels={Trivago, Spotify, Microsoft},
legend style={at={(1.25,0.50)}, anchor=west},
]
\addplot[thin, fill=cyan, postaction={}, area legend] table [y=bounded, x expr=\coordindex, col sep=comma] {csv/bounds.csv};
\addplot[thin, fill=red, postaction={}, area legend] table [y=left-bounded, x expr=\coordindex, col sep=comma] {csv/bounds.csv};
\addplot[thin, fill=blue, postaction={}, area legend] table [y=right-bounded, x expr=\coordindex, col sep=comma] {csv/bounds.csv};
\addplot[thin, fill=orange, postaction={ }, area legend] table [y=unbounded, x expr=\coordindex, col sep=comma] {csv/bounds.csv};
\legend{bounded, left-bounded, right-bounded, unbounded}
\end{axis}
\end{tikzpicture}
\caption{Share of bounded, left-bounded, right-bounded, and unbounded communication channels among all cases. }
\label{tab:bound}
\end{figure*}
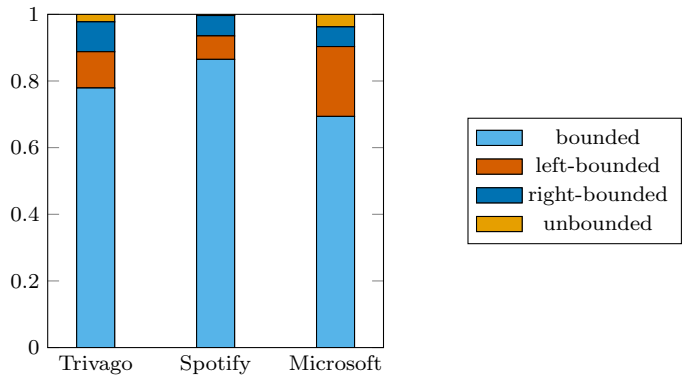

In particular, in communication channels (code reviews) that are right-bounded or unbound, we may miss participants who contributed to the discussion and, therefore, can spread information. Left-bounded and bounded communication channels do not contribute to this uncertainty since we know all participants within the observation window. In \Cref{tab:bound}, we see the distributions of the communications-channel bounds.

\begin{figure*}
\centering
\pgfplotsset{horizons axis/.style={
	axis x line*=bottom,
	axis y line*=left,
	axis on top,
	width=10cm,
	height=5cm,
	enlargelimits=false,
	xtick={0, 7, 14, 21, 28},
	xticklabels={},
	xmin=0,
	xmax=28,
	ymin=0,
	xtick align=outside,
	ytick align=outside,
	every tick/.style={black},
	ylabel absolute,
	ylabel style={yshift=1em},
}}
\begin{tikzpicture}
\matrix{ %
\begin{axis}[%
	horizons axis,
	ylabel=Trivago,
	ymax=360,
]
\addplot graphics[xmin=0, ymin=0, xmax=28, ymax=360] {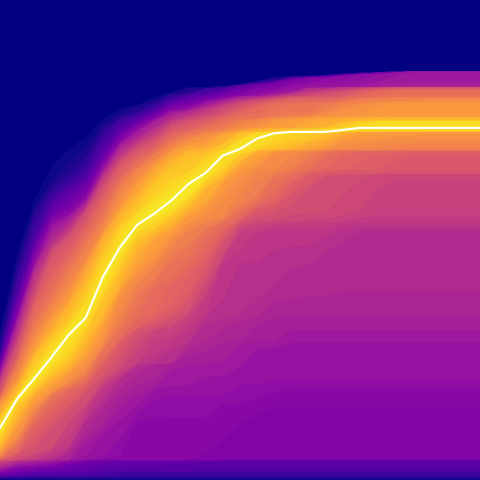};
\end{axis}
\begin{axis}[
	horizons axis,
	hide x axis,
	axis y line*=right,
	ymin=0, ymax=1,
]
\end{axis}\\
\begin{axis}[%
	horizons axis,
	ylabel=Spotify,
	ymin=0, ymax=1739,
]
\addplot graphics[xmin=0, ymin=0, xmax=28, ymax=1739] {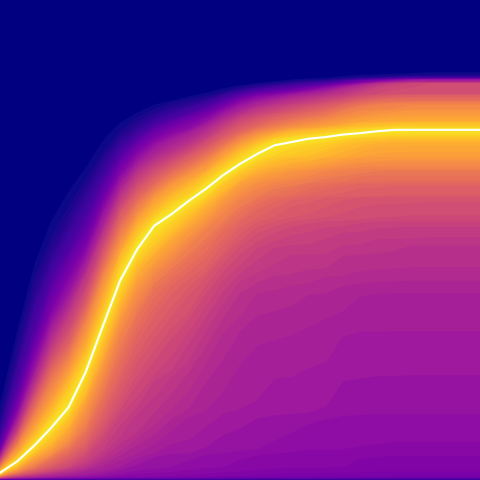};
\end{axis}
\begin{axis}[
	horizons axis,
	hide x axis,
	axis y line*=right,
	ymin=0, ymax=1,
]
\end{axis}\\
\begin{axis}[%
	horizons axis,
	ylabel=Microsoft,
	ymin=0, ymax= 37103,
	scaled y ticks=false,
	extra x ticks={3.5, 10.5, 17.5, 24.5},
	extra x tick labels={Week 1, Week 2, Week 3, Week 4},
	extra x tick style={tick style={draw=none}},
]
\addplot graphics[xmin=0, ymin=0, xmax=28, ymax=37103] {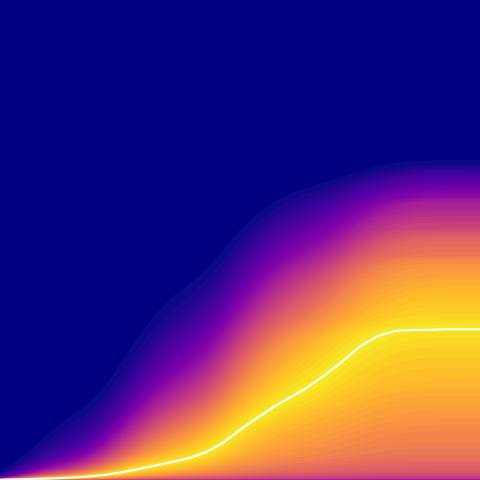};
\end{axis}
\begin{axis}[
	horizons axis,
	hide x axis,
	axis y line*=right,
	ymin=0, ymax=1,
]
\end{axis}\\
};
\end{tikzpicture}
\caption{The information diffusion range absolute (left $y$-axis) and relative to the network size (right $y$-axis) distribution per participant in the observation window of four weeks ($x$-axis). The distribution is presented as a color-coded intra-percentile range with the median in white. }
\label{fig:coverage_over_time}
\end{figure*}

Although the observation windows of four weeks are arbitrary and, therefore, all distances longer than four weeks are not captured, we consider the observation window as sufficiently large enough to capture the information diffusion through code review: All code review networks reached a plateau regarding how far information can spread through code review. \Cref{fig:coverage_over_time} is a comprehensive overview of our simulation results depicting the distributions of reachable code review participants over time as color-coded inter-percentile ranges for all three code review systems. 

We observe that the code review systems at Trivago and Spotify reach a plateau after two, and the code review system at Microsoft after three weeks on how far information can spread. That means a larger observation window and, therefore, longer topological and temporal distances would likely not significantly impact the information diffusion and the number of reachable participants. 

Not only the size of the observation window but also its positioning in time can affect our results. Larger discontinuities (e.g., holidays for large parts of the staff, vacation seasons over summer) and external interferences (e.g., the pandemic, large-scale outages of the development infrastructure, etc.) with the software development will affect code review, and, thus, impact our results. We void such noise that would have impacted our results by carefully selecting pre-pandemic observation windows with no significant discontinuities by public holidays such as Christmas (see \Cref{sec:data_collection}). Therefore, we believe that the positioning of the timeframes in time for code review as a continuous endeavor has no significant influence on the results. However, we are not able to provide empirical evidence to validate this claim (i.e., the observation window covers a typical, presentative month) beyond the assessment of our industry partners when selecting the timeframe.

\subsection{Generalization}

The generalization of our results highly depends on our sampling strategy. In general, our study is affected by an availability bias: Companies are hesitant to share code review data since the code review system---as any communication tool---may contain confidential and personally identifiable information of their developers. However, we used a maximum variation sampling to select suitable code review systems in an industrial context and aimed for representativeness on code review system size and tooling. 

The size classification of code review systems is arbitrary and relative to our sample. On the lower bound, we classified a code review system with \num{364} participants to be small, although there are code review systems that are significantly smaller. On the upper bound, however, our sample includes arguably one of the largest code review systems nowadays with more than \num{37103} participants. 

Our sampling contains three code review tools: GitHub (via pull/merge requests) at Spotify, BitBucket (via merge requests) at Trivago, and CodeFlow as an internally developed tool at Microsoft. This, however, does not cover the tool landscape extensively: In particular, we miss \emph{Gerrit}, \emph{Phabricator}, and \emph{Gitlab} from the broadly available tools. We believe that missing Phabricator and Gitlab does not introduce a bias: The future of Phabricator is unknown since the company running the developed, \emph{Phacility}, ended all operations \citep{Phabricator} and the code review (via merge requests) in Gitlab is, in our opinion, equivalent to that in GitHub. However, Gerrit, as a popular and dedicated code review tool with its voting system, could lead to other communication networks and, therefore, to different information diffusion results. 

We explicitly excluded the practices as sampling dimensions although they have a direct impact on the resulting communication network: Code review guidelines, code ownership, or other selection criteria define who is invited to and has a say in a code review discussion and, therefore, prescribe and limit the available communication channels for the information sharing. Those guidelines vary among companies but also within them. 


\section{Conclusion}\label{sec:conclusion}

With this study, we make a first quantitative step towards understanding and evaluating code review as a communication network at scale. Through our information diffusion simulation based on communication networks emerging from code review systems at Microsoft, Spotify, and Trivago, we found an upper bound of information diffusion in code review: 

On average (median), information can spread between \qty{72}{\percent} and \qty{85}{\percent} of all participants for small and mid-sized code review systems or between \qty{30}{\percent} and \qty{71}{\percent} of all participants for large code review systems (Microsoft); which corresponds to an absolute range between \num{11645} and \num{26216} code review participants. This describes the upper bound on how far information can spread in code review

The average (median) distance between two code review participants is shorter than three code reviews or five days at small code review systems (Trivago). The average distance in mid-sized (Spotify) and large (Microsoft) code review systems ranges between four and seven code reviews or seven and fourteen days---considering the sizes on an exponential-like scale a significant finding. This describes the upper bound on how fast information can spread in code review. 

Our findings align with findings from the prior qualitative research: All five relevant prior studies reported information exchange among code review participants as a key expectation towards code review. Our findings (indicating that the communication network that emerges from code review is capable of diffusing information fast and far) corroborate the qualitatively reported information exchange as an expectation towards code review, which we consider the foundation for all other reported and expected benefits of code review. Although we argue that given the sheer magnitude of information diffusion possible through code, information must cross organizational boundaries and, therefore, code review enables collaboration in companies at scale, future work is required to investigate and establish a thorough connection between our work and an improved collaboration at scale, another qualitatively reported expectation towards code review. This applies to code-related expected improvements, too. 

Although our sample of code review systems limits the generalizability of our findings, we conclude for researchers and practitioners alike:

\begin{itemize}
\item {\emph{The larger the better.}} Because code review is a communication network that can scale with the information diffusion among its participants, companies may unify and centralize their code review systems---independently of (monolithic) code repositories \citep{Potvin2016}. 
\item {\emph{Tooling is secondary.}} We did not find any evidence that the choice of tooling plays a crucial role in information diffusion through code review.
\item {\emph{The role of bots rethought.}} Although bots can provide assistance for selecting and informing a set of code review participants optimal with respect to information diffusion, bots tend to introduce noise in the communication channels \citep{Wessel2021} that may slow down the communication through code review. The promising work by \cite{Roeseler2023} can lay the foundation for understanding the impact of bots on communication, coordination, and collaboration. 
\end{itemize}

Our comprehensive replication package enables researchers to fully reproduce our results and replicate our study using other code review systems to parametrize our simulation model. 

The need for replication applies in particular to open source. Not only because \cite{Bosu2017} already reported a significantly large difference in the primary motivation for code review in an industrial and open-source context, we believe that findings from an industrial or open-source context are not easily transferrable: The mechanics and incentives in open source differ, and so do the organizational structure, liability, and commitment \citep{Barcomb2020}. 

We also invite to enhance our simulation. In particular, implementing diffusion probabilities could broaden our understanding of information diffusion beyond the best-case assumptions. 

So far, we have explicitly excluded the underlying code review practices as a sampling dimension. These practices most likely significantly impact the resulting communication networks emerging from code review and, therefore, the information diffusion in code review. Future work could map practices to information diffusion to indicate the reasonable cost-benefit ratio of code review. 

In this study, we focused on the upper bound of information diffusion and answered the research questions regarding how far and how fast information can spread through code review. Our research design does not aim to investigate how fast and how far information actually spreads through code review; it remains an estimation of the upper bound of information diffusion in code review. In future work, we will measure (rather than simulate) the actual information diffusion through code review. Therefore, we will develop a measurement system to follow the traces in the communication networks that emerge from code review. Code review tools like GitHub provide a foundation for those investigations.

\section*{Code and Data Availability}

The code and data that support the findings of this study are openly available on Zenodo via \cite{Dorner2023data} and \cite{Dorner2023software}.

\section*{Acknowledgement}

We thank Andreas Bauer for his valuable feedback on the technical aspects of the simulation. We are very grateful for the support from our industry partners, in particular, from Andy Grunwald, Simon Brüggen, and Marcin Floryan. We also thank the anonymous reviewers for their careful reading of our manuscript and their many insightful comments and helpful suggestions, and Daniel Graziotin from the Open Science Board at EMSE for his prompt and profound feedback on our replication package. This work was supported by the KKS Foundation through the SERT Project (Research Profile Grant 2018/010) at Blekinge Institute of Technology.

\section*{Contribution Statement} 

\textbf{Michael Dorner:} Conceptualization, Data curation, Methodology, Software, Formal analysis, Investigation, Writing - Original Draft, Visualization, Supervision
\textbf{Daniel Mendez:} Conceptualization, Funding acquisition, Writing - review \& editing, Supervision
\textbf{Krzysztof Wnuk:} Conceptualization, Writing - review \& editing
\textbf{Ehsan Zabardast:} Data curation, Writing - review \& editing
\textbf{Jacek Czerwonka:} Data curation, Writing - review \& editing
\textbf{Andreas Bauer:} Software, Validation
\textbf{Andy Grunwald:} Data curation
\textbf{Simon Brüggen:} Data curation

\section*{Conflict of Interest}

The authors declared that they have no conflict of interest.

\bibliographystyle{spbasic} 
\bibliography{references}

\end{document}

%% file: tikz-hypergraph.tex
\def\rotateclockwise#1{
  \newdimen\xrw
  \pgfextractx{\xrw}{#1}
  \newdimen\yrw
  \pgfextracty{\yrw}{#1}
  \pgfpoint{\yrw}{-\xrw}
}

\def\rotatecounterclockwise#1{
  \newdimen\xrcw
  \pgfextractx{\xrcw}{#1}
  \newdimen\yrcw
  \pgfextracty{\yrcw}{#1}
  \pgfpoint{-\yrcw}{\xrcw}
}

\def\outsidespacerpgfclockwise#1#2#3{
  \pgfpointscale{#3}{
    \rotateclockwise{
      \pgfpointnormalised{
        \pgfpointdiff{#1}{#2}}}}
}

\def\outsidespacerpgfcounterclockwise#1#2#3{
  \pgfpointscale{#3}{
    \rotatecounterclockwise{
      \pgfpointnormalised{
        \pgfpointdiff{#1}{#2}}}}
}

\def\outsidepgfclockwise#1#2#3{
  \pgfpointadd{#2}{\outsidespacerpgfclockwise{#1}{#2}{#3}}
}

\def\outsidepgfcounterclockwise#1#2#3{
  \pgfpointadd{#2}{\outsidespacerpgfcounterclockwise{#1}{#2}{#3}}
}

\def\outside#1#2#3{
  ($ (#2) ! #3 ! -90 : (#1) $)
}

\def\cornerpgf#1#2#3#4{
  \pgfextra{
    \pgfmathanglebetweenpoints{#2}{\outsidepgfcounterclockwise{#1}{#2}{#4}}
    \let\anglea\pgfmathresult
    \let\startangle\pgfmathresult

    \pgfmathanglebetweenpoints{#2}{\outsidepgfclockwise{#3}{#2}{#4}}
    \pgfmathparse{\pgfmathresult - \anglea}
    \pgfmathroundto{\pgfmathresult}
    \let\arcangle\pgfmathresult
    \ifthenelse{180=\arcangle \or 180<\arcangle}{
      \pgfmathparse{-360 + \arcangle}}{
      \pgfmathparse{\arcangle}}
    \let\deltaangle\pgfmathresult

    \newdimen\x
    \pgfextractx{\x}{\outsidepgfcounterclockwise{#1}{#2}{#4}}
    \newdimen\y
    \pgfextracty{\y}{\outsidepgfcounterclockwise{#1}{#2}{#4}}
  }
  -- (\x,\y) arc [start angle=\startangle, delta angle=\deltaangle, radius=#4]
}

\def\corner#1#2#3#4{
  \cornerpgf{\pgfpointanchor{#1}{center}}{\pgfpointanchor{#2}{center}}{\pgfpointanchor{#3}{center}}{#4}
}

\def\hedgem#1#2#3#4{
  
  \outside{#1}{#2}{#4}
  \pgfextra{
    \def\hgnodea{#1}
    \def\hgnodeb{#2}
  }
  foreach \c in {#3} {
    \corner{\hgnodea}{\hgnodeb}{\c}{#4}
    \pgfextra{
      \global\let\hgnodea\hgnodeb
      \global\let\hgnodeb\c
    }
  }
  \corner{\hgnodea}{\hgnodeb}{#1}{#4}
  \corner{\hgnodeb}{#1}{#2}{#4}
  -- cycle
}

\def\hedgeii#1#2#3{
  \hedgem{#1}{#2}{}{#3}
}

%% file: tex/normalized_ranges.tex
\begin{tabular}{@{}llllll@{}}
\toprule
 & \num{0.70} & \num{0.50} & \num{0.30} & \num{0.10} & max \\
\midrule
Trivago & \numrange{0.67}{0.85} & \numrange{0.72}{0.85} & \numrange{0.81}{0.85} & \numrange{0.83}{0.85} & \num{0.85} \\
Spotify & \numrange{0.58}{0.85} & \numrange{0.72}{0.85} & \numrange{0.79}{0.85} & \numrange{0.83}{0.85} & \num{0.85} \\
Microsoft & \numrange{0.01}{0.71} & \numrange{0.30}{0.71} & \numrange{0.50}{0.71} & \numrange{0.61}{0.71} & \num{0.71} \\
\bottomrule
\end{tabular}

%% file: tex/ranges.tex
\begin{tabular}{@{}llllll@{}}
\toprule
 & \num{0.70} & \num{0.50} & \num{0.30} & \num{0.10} & max \\
\midrule
Trivago & \numrange{245}{310} & \numrange{266}{310} & \numrange{296}{310} & \numrange{309}{310} & \num{310} \\
Spotify & \numrange{1026}{1472} & \numrange{1260}{1472} & \numrange{1386}{1472} & \numrange{1447}{1472} & \num{1472} \\
Microsoft & \numrange{808}{26216} & \numrange{11645}{26216} & \numrange{18887}{26216} & \numrange{22983}{26216} & \num{26216} \\
\bottomrule
\end{tabular}